\documentclass[%
preprint,
showpacs,preprintnumbers,
nofootinbib,
nobibnotes,
 amsmath,amssymb,
prd,
floatfix
]{revtex4}

\usepackage{graphicx}
\usepackage{dcolumn}
\usepackage{bm}

\usepackage[
letterpaper,
top=1in, bottom=1in,left=1in,right=1in
]{geometry}

\usepackage[usenames]{color}

\usepackage{subfigure}
\usepackage{graphics}
\usepackage{amssymb}
\usepackage{rotating}

\newcommand{\bea}{\begin{eqnarray}}
\newcommand{\eea}{\end{eqnarray}}

\def\beq{\begin{equation}}
\def\eeq{\end{equation}}

\def\beqn{\begin{eqnarray}}
\def\eeqn{\end{eqnarray}}
\def\ba{\begin{eqnarray}}
\def\ea{\end{eqnarray}}

\newcommand{\be}{\begin{equation}}
\newcommand{\beqa}{\begin{eqnarray}}
\newcommand{\eeqa}{\end{eqnarray}}

\newcommand{\ee}{\end{equation}}

\begin{document}

\preprint{ANL-HEP-PR-10-19, SMU-HEP-10-09
}

\title{Constraints on color-octet fermions from a global parton distribution analysis}

\author{Edmond L.~Berger$^1$, 
Marco Guzzi$^2$, Hung-Liang~Lai$^3$, Pavel M. Nadolsky$^2$, and 
Fredrick I. Olness$^2$}

\affiliation{
$^1$ High Energy Physics Division, Argonne National Laboratory, Argonne, Illinois 60439, USA\\
$^2$ Department of Physics, Southern Methodist University, Dallas Texas 75275, USA\\
$^3$Taipei Municipal University of Education, Taipei, Taiwan
}%

\date{\today}

\begin{abstract}

We report a parton distribution function (PDF) analysis of a complete 
set of hadron scattering data, in which a color-octet fermion 
(such as a gluino of supersymmetry) is incorporated as an extra parton constituent along with 
the usual standard model constituents.    The data set includes the 
most up-to-date results from deep-inelastic scattering and from jet 
production in hadron collisions.  Another feature is the inclusion in 
the fit of data from determinations  of the strong coupling $\alpha_s(Q)$ 
at large and small values of the hard scale $Q$.  Our motivation is 
to determine the extent to which the global PDF analysis may 
provide constraints on the new fermion, as a function of its mass and 
$\alpha_s(M_Z)$, independent of assumptions such as the mechanism 
of gluino decays.  Based on this analysis, we find that gluino masses as 
low as  30 to 50 GeV may be compatible with the current hadronic data. 
Gluino masses below 15 GeV (25 GeV) are excluded if $\alpha_s(M_Z)$ 
varies freely (is equal to 0.118).  At the outset, 
stronger constraints had been anticipated from jet production 
cross sections, but experimental systematic uncertainties, particularly 
in normalization, reduce the discriminating power of these data.   
\end{abstract}

\pacs{
13.87.-a,       
14.80.Ly,       
13.85.Qk       
}

\maketitle


\section{Introduction \label{sec:level1}
}

Heavy color-octet particles are postulated in theories of
beyond-the-standard-model (BSM) phenomena, including supersymmetry
(SUSY)~\cite{Haber:1984rc}, universal extra
dimensions~\cite{Appelquist:2000nn},
Randall-Sundrum~\cite{Randall:1999ee}, and Little Higgs
models~\cite{ArkaniHamed:2001nc}.
Direct searches for such states are usually guided by aspects of the
production and decay dynamics in the particular BSM approach.
Analyses of search data have so far produced various bounds on the
masses of the states, often conditioned by model-dependent
assumptions~\cite{Alitti:1989ux,Abazov:2006bj,Abazov:2007ht,:2007ww,Abazov:2009rj,Aaltonen:2009fm,Aaltonen:2008rv,Abulencia:2005us,Abulencia:2006kk,Alwall:2008ve,Berger:2008cq}.
Different constraints,  such as the SUSY gluino mass bounds
$m_{\tilde{g}} > 26.9 $ GeV \cite{Heister:2003hc} and 51 GeV \cite{Kaplan:2008pt}
at 95\% confidence level (C.L.), are based on the analysis of LEP
event shapes in soft-collinear effective theory and other quantum chromodynamics (QCD) 
resummation formalisms.   
Constraints such as these may depend on theoretical modeling 
of nonperturbative hadronization and the matching of 
hard-scattering and resummed contributions,  similar to the determination of 
$\alpha_s(M_Z)$ from LEP data in QCD~\cite{Becher:2008cf, Davison:2008vx, Dissertori:2009ik, Abbate:2010vw, Abbate:2010xh}.   
In a previous publication~\cite{Berger:2004mj}, we examine the possibility that 
a global analysis of hadron data, within the framework of parton distribution function (PDF)
determinations, can be used to derive constraints on the existence and masses of color-octet 
fermions, independently of other information on such states.   Global analysis has 
discriminating power for several reasons: one is that new colored states modify the evolution with hard 
scale $Q$ of the strong coupling strength $\alpha_s(Q)$.  Second, in perturbative 
QCD, the coupling of a color-octet fermion to
quarks and gluons alters the set of evolution equations that governs
the behavior  of all parton distribution functions, thus affecting many hadron scattering
cross sections.  Moreover, production of the color-octet states will affect relevant observables, such as 
jet rates, whose cross sections are included in the global fits.  

The specific case of a gluino from supersymmetry is included as an extra degree of freedom 
in our earlier work~\cite{Berger:2004mj}.   We refer to the PDFs obtained in that publication 
as ``SUSY PDFs'', although our analysis is applicable to a broader class 
of standard model (SM) extensions.  In Ref.~\cite{Berger:2004mj}, a lower bound on the gluino 
mass $m_{\tilde g}$ is obtained in terms of an assumed value of $\alpha_s(M_Z)$ at $Z$ 
boson mass $M_Z$.   For the then standard model world-average value
of $\alpha_s(M_Z) = 0.118$, gluinos lighter than 12~GeV were shown to be
disfavored,  whereas the lower bound was relaxed to less than
10~GeV (less than 2~GeV) when $\alpha_s(M_Z)$ was increased above 0.120 (0.127).

In this paper, we use new hadron scattering data incorporated in the
next-to-leading order (NLO)  CT10 general-purpose PDF analysis~\cite{CT10}, along 
with a new approach for incorporating the variation of $\alpha_s(Q)$ into PDF
determinations~\cite{Lai:2010nw}, to obtain improved bounds on the mass of a relatively light gluino.  
The essential new elements are these:  
\begin{itemize}
\item
{\bf New Tevatron jet data~\cite{Aaltonen:2008eq,Abulencia:2007ez,2008hua} and combined DIS data~\cite{2009wt} 
from HERA. }  In a global QCD analysis, the presence of light gluinos is revealed primarily
by modifications of $\alpha_s(M_Z)$, the gluon PDFs, 
and the charm and bottom quark PDFs, generated radiatively above the respective 
heavy-quark thresholds. 
We include the latest hadronic scattering data sensitive
to such modifications.  The most stringent constraints on
the gluon PDF are imposed by electron-proton deep inelastic
scattering (DIS) data at $x<0.1$ and single-inclusive jet production data from 
the Tevatron $p\bar p$ collider at $x\gtrsim 0.1$.   The study reported here incorporates 
up-to-date information from the combined H1 Collaboration and ZEUS 
Collaboration data on deep inelastic 
scattering  at HERA-1~\cite{2009wt}, as well as single-inclusive jet data 
from the Tevatron Run-II analyses~\cite{Aaltonen:2008eq,Abulencia:2007ez,2008hua}.
Hard scattering contributions of massive gluinos, with full dependence
on the gluino's mass, are included in the jet production cross
sections we use, the only process we examine where these contributions are
large enough to be relevant at NLO accuracy.
\item
{\bf Floating $\alpha_s(M_Z)$.} Our fits are performed by treating
$\alpha_s(M_Z)$ at the mass $M_Z$ of the $Z$ boson 
as a variable parameter of the standard model.  
We constrain $\alpha_s(M_Z)$ by requiring that the fitted 
$\alpha_s(Q)$ agree with its direct determinations at low energy
scales ($Q<10$ GeV) and at $Q=M_Z$, within the quoted uncertainties of these measurements. 
Virtual gluino contributions result in a slower evolution of the QCD coupling 
strength $\alpha_s(Q)$ at scales $Q$ above the gluino mass
threshold.  By including data that constrain $\alpha_s$ at low
and high $Q$ scales, we effectively probe for deviations from pure QCD. 
We find, in particular, that the value of $\alpha_s(M_Z)=0.123\pm 0.004$ 
derived in some analyses of LEP
event shapes~\cite{Dissertori:2009ik} can be accommodated if gluinos have 
mass of about 50 GeV. 
\end{itemize}

The remainder of this paper is organized as follows.  In Sec.~\ref{sec:Overview}, 
we describe the role of  new color-octet fermions in a global QCD analysis.  
The incorporation of data on $\alpha_s(Q)$ within the global fits is 
discussed in Sec.~\ref{sec:alpha}, where we also present the values of 
$\alpha_s(Q)$ at high and low $Q$ used in our fits.   Our 
simultaneous global fit to hadronic scattering data and $\alpha_s(Q)$ 
is described in Sec.~\ref{sec:results}, where we also examine  
the effects of an additional gluino degree of freedom on the PDFs.   
We present figures that show the relative
magnitudes of the PDFs and the variation of their momentum fractions
with gluino mass and hard scale. 

Section~\ref{sec:compwdata} contains the results of our detailed comparison 
with data.   We present figures that show the variation of the values of $\chi^2$ 
in the global analyses, as a function of gluino mass, for both floating and fixed 
$\alpha_s(M_Z)$.    Section~\ref{sec:compwdata} also 
includes the comparison of our  calculated cross sections with jet data from 
the Tevatron collider and a discussion of the systematic uncertainties that 
limit the constraining power of these data.
The sensitivity of jet cross sections at the LHC to the presence of gluinos is 
examined in Sec.~\ref{sec:LHC}.  Our conclusions are presented in 
Sec.~\ref{sec:Conclusion}.    The appendices contain an analytic expression 
for the evolution of the strong 
coupling $\alpha_s(Q)$ in terms of the SM and SUSY degrees of freedom;  
expressions for the contributions of massive gluinos 
to the jet production cross sections; and  
parton-parton luminosity functions for various  
combinations of  SM partons and gluinos.  

Based on our analysis, we conclude that gluino masses as low as  30 to 50 GeV 
may be compatible with the current hadronic data, depending on the value of 
$\alpha_s(M_Z)$.   For a floating $\alpha_s(M_Z)$, gluinos lighter than 15 GeV are 
excluded.  For an assumed fixed value 
$\alpha_s(M_Z)=0.118$,  the world average value used in many 
phenomenological analyses, gluinos lighter than 25 GeV are disfavored.     

We acknowledge that a gluino as light as $\sim 50$~GeV is not typical in 
phenomenological models of SUSY breaking, nor of the results of 
experimental direct search analyses based on specific models of 
SUSY breaking and assumptions about mass relationships among SUSY 
states~\cite{Berger:2008cq}.  As long as the SUSY neutralino $\widetilde \chi^0$ is lighter 
than the gluino, the typical decay process for a light gluino is 
$\widetilde g \rightarrow q \bar q \widetilde \chi^0$, where $q$ stands for a SM quark. 
Missing energy would signal the presence of a neutralino.  However, for a small 
mass splitting $m_{\widetilde g} - m_{\widetilde \chi^0}$, the 
gluino's decay into missing 
energy and soft quark  jets would be undetected.   The analysis reported here is 
complementary to other approaches for bounding the gluino mass, and it 
is in some respects more general in that we make no assumptions about the 
gluino decay.  

Precise determination of $\alpha_s(M_Z)$ and proton PDFs 
are essential ingredients for obtaining reliable predictions from perturbative 
QCD calculations.   Such calculations are key for the general physics program 
and for new physics searches at the CERN Large Hadron Collider (LHC) and 
Fermilab Tevatron  collider.  As we show here, these ingredients themselves 
may be affected by non-SM contributions, at all values of the momentum
fraction $x$, as a result of the global interconnections in PDF analyses.   
The determination of the QCD coupling $\alpha_s$ and of the gluon PDF from 
the Tevatron or LHC single-inclusive jet data, such as in recent Tevatron Run-2 
measurements~\cite{Aaltonen:2008eq,Abulencia:2007ez,2008hua,Abazov:2009nc}, 
may be sensitive to scattering of color-octet fermions in
the ways discussed in Sec.~\ref{sec:LHC}.   As a result of our work, we determine new
sets of  PDFs  that include a relatively light gluino as a hadron constituent.  

\section{Color-octet fermions in a global QCD analysis\label{sec:Overview}}

Under well-defined conditions, a relatively light strongly-interacting
fundamental particle may be treated as a constituent of the colliding
hadrons. It will share the momentum of the parent hadron with the
standard model quark, antiquark, and gluon partners. The experimental
consequences of this picture become evident when the parent hadron
is probed at a sufficiently large hard scale. For example, the charm
quark $c$ and bottom quark $b$ are treated appropriately as partonic
constituents of hadrons when the characteristic energy scale $Q$
exceeds the mass of the heavy quark $m_{q}$. Likewise, when $Q$
greatly exceeds the mass of a new strongly-interacting particle, this object
must also be incorporated as a hadronic constituent.
We refer to Ref.~\cite{Berger:2004mj} for an exposition of the PDF
analysis in which a gluino is included as an additional partonic degree of freedom.   

As in Ref.~\cite{Berger:2004mj}, we take the gluino as the only colored non-SM  degree 
of freedom that needs to be considered.   In some models of SUSY breaking, such as 
split supersymmetry~\cite{ArkaniHamed:2004fb,Giudice:2004tc},  the squarks are much heavier than the gluinos, 
and therefore could be omitted from our PDF  analysis.  Moreover, as illustrated 
in Eq.~(A3) of Appendix A,  color-octet spin-1/2 fermions 
have a greater impact on the evolution of the strong coupling $\alpha_s$ than color triplet scalars, such as squarks.\footnote{While bottom squarks
($\tilde{b}$) can be relatively light in some models~\cite{Berger:2000mp}, 
their contribution to DIS and other relevant 
cross sections can be neglected, cf.~Ref.~\cite{Berger:2004mj}.} 

The presence of a light gluino $\tilde{g}$ modifies the PDF global
analysis in three ways.%

\begin{enumerate}
\item The gluino changes the evolution of the strong coupling strength $\alpha_{s}(Q)$, as the scale $Q$ is varied. This influence is implemented in our
results, and we provide details on the running of $\alpha_{s}(Q)$
in Appendix~\ref{app:alphas}.   The constraints on the gluino mass from our global analysis depend 
significantly on the value of the strong coupling strength $\alpha_{s}(M_{Z})$. 
\item The gluino provides an additional partonic degree of freedom that
shares in the nucleon's momentum. It alters the coupled set of
Dokshitzer-Gribov-Lipatov-Altarelli-Parisi equations
that govern the evolution of the parton distributions,
\begin{small}
\begin{eqnarray}
 &  & Q ^{2}\frac{d}{dQ ^{2}}\left(\begin{array}{c}
 \Sigma (x,Q )\\
 g(x,Q )\\
 \tilde{g}(x,Q )\end{array}
\right)=\frac{\alpha _{s}(Q)}{2\pi }\, \, \times \nonumber \\
 & & \int _{x}^{1}\frac{dy}{y}\left(\begin{array}{ccc}
 P^{NLO}_{\Sigma \Sigma }\bigl (x/y\bigr ) & P_{\Sigma g}^{NLO}\bigl (x/y\bigr ) & P_{\Sigma \tilde{g}}^{LO}\bigl (x/y\bigr )\\
 P^{NLO}_{g\Sigma }\bigl (x/y\bigr ) & P^{NLO}_{gg}\bigl (x/y\bigr ) & P^{LO}_{g\tilde{g}}\bigl (x/y\bigr )\\
 P^{LO}_{\tilde{g}\Sigma }\bigl (x/y\bigr ) & P^{LO}_{\tilde{g}g}\bigl (x/y\bigr ) & P^{LO}_{\tilde{g}\tilde{g}}\bigl (x/y\bigr )\end{array}
\right)\left(\begin{array}{c}
 \Sigma (y,Q )\\
 g(y,Q )\\
 \tilde{g}(y,Q )\end{array}
\right)\, ;
\label{SAP}
\nonumber
\end{eqnarray}
\end{small}
\begin{eqnarray}
\Sigma (x,Q ) = \sum_{i=u,d,s,...} (q_i(x,Q ) + {\bar q_i}(x,Q )).  
\end{eqnarray}
Here $\Sigma (x,Q )$, $g(x,Q )$, and $\tilde{g}(x,Q )$
are the singlet quark, gluon, and gluino distributions, respectively; 
$q_i(x,Q )$ and ${\bar q_i}(x,Q )$ are the quark and antiquark 
distributions for a flavor $i$.  
The previous analysis~\cite{Berger:2004mj} shows that the gluino's 
contribution is small in the momentum fraction range $x> 10^{-5}$, 
$\tilde{g}(x,Q)\ll g(x,Q)$, and $\tilde{g}(x,Q)\ll q(x,Q)$.  NLO variations in the 
relevant SUSY cross sections are  small and comparable in size to 
variations associated with next-to-next-to-leading order (NNLO) 
SM contributions. Therefore,
the leading-order (LO) approximation for the splitting functions and 
hard scattering amplitudes of SUSY terms is numerically adequate, when combined
with NLO expressions for SM contributions. 

\item At energies above its mass threshold, a color-octet fermion contributes to
hard scattering processes as an incident parton and/or as a produced
particle. However, as argued in Ref.~\cite{Berger:2004mj}, in the
absence of light squarks, gluino hard-scattering contributions to DIS and
Drell-Yan process are of next-to-next-to-leading order and negligible
in the current study. At the same time, the hard-scattering gluino
terms contribute at the LO in single-inclusive jet production, 
so that it is essential that we include the
gluino in the corresponding hard scattering matrix elements of jet
cross sections. 

The $2\rightarrow 2$ hard scattering contributions with two gluinos 
in the initial or final states 
are illustrated in Fig.~\ref{feyn_gluino}. We assume that the masses
of the squarks are large enough that diagrams containing a squark
propagator are negligible.
\begin{figure}[tb]
\includegraphics[width=11cm, angle=0]{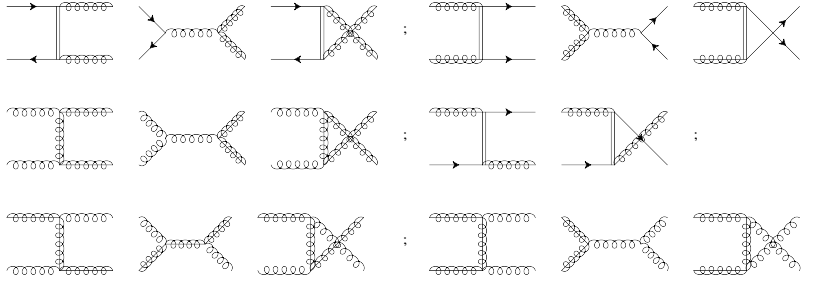}
\caption{\label{feyn_gluino} LO scattering diagrams with gluinos in the initial or final state.  
The double lines stand for the squark exchange contributions that we neglect in our 
approximation.}
\end{figure}
The remaining SUSY diagrams can be evaluated in the the S-ACOT
factorization scheme~\cite{Collins:1998rz,Kramer:2000hn}, in order to
simplify treatment of the gluino mass dependence. In this scheme,
gluino mass terms are retained in diagrams with two final-state
gluinos in the subprocesses $gg
\rightarrow\tilde{g}\tilde{g}$ and $q\bar q
\rightarrow\tilde{g}\tilde{g}$. Explicit scattering amplitudes in
these channels are documented in 
Eqs.~(\ref{qqb_ginogino}) and (\ref{gg_ginogino}) of Appendix~\ref{app:jet}.
Massless amplitudes are used for the
remaining $2\rightarrow 2$ hard scattering subprocesses, 
in which one or two gluinos are present in the initial state, and whose 
contributions are proportional to the gluino PDF $\tilde{g}(x,Q)$.
This arrangement captures the full gluino mass dependence, while 
including the mass terms only in the essential scattering amplitudes.
\end{enumerate}

\section{QCD coupling strength as a fitting parameter\label{sec:alpha}}

Since the range of $m_{\tilde g}$ values
allowed by the global fits depends strongly 
on the assumed value of $\alpha_s(M_Z)$, we do a simultaneous 
fit to hadronic data and to data on  $\alpha_s(Q)$ in this work.   
A judicious choice is required 
therefore of the set of data on  $\alpha_s(Q)$.    

Our approach is to fit the global set of data using $\alpha_s(Q)$ as a
floating parameter, constraining it with additional data on $\alpha_s(Q)$ 
measurements at $Q<10$ GeV 
(i.e., in the range where gluino contributions are excluded by the
previous analysis), and at $Q=M_Z$ (in $e^+e^-$
hadroproduction at LEP). This approach is similar to the floating
$\alpha_s(M_Z)$ fit in Ref.~\cite{Lai:2010nw}.  
However, we constrain $\alpha_s(Q)$ at
two distinct $Q$ values, to probe for deviations of its running
from the SM prediction. 

\subsection{Low-energy constraints \label{sec:LowEnergy}}
The QCD coupling constraint at low $Q=5$ GeV, 
\be 
\alpha_s(Q=5\mbox{ GeV})=0.213 \pm 0.002,
\label{alphas:lowQ}
\ee
is obtained as a weighted average of three precise determinations of
$\alpha_s$ at comparable energies:
\ba
&&\alpha_s(Q=5\mbox{ GeV}) = 0.219 \pm 0.006 \hspace{.5cm} \textrm{from $\tau$ decays},\\
&&\alpha_s(Q=5\mbox{ GeV}) = 0.214 \pm 0.003 \hspace{.5cm} \textrm{from heavy quarkonia},\\
&&\alpha_s(Q=5\mbox{ GeV}) = 0.209 \pm 0.004 \hspace{.5cm} \textrm{from lattice QCD}. 
\ea
These values are reconstructed by QCD evolution to the common scale $Q=5$ GeV of the published
$\alpha_s$ values provided at different energy scales, 
\ba
&&(\alpha_s)_\tau = 0.330\pm 0.014 \mbox{ at } m_{\tau}=1.77 \mbox{ GeV},\\
&&(\alpha_s)_{Q\bar{Q}} = 0.1923 \pm 0.0024 \mbox{ at } M_{Q\bar{Q}}=7.5 \mbox{ GeV},\\
&&(\alpha_s)_{\rm lattice} = 0.1170 \pm 0.0012 \mbox{ at } M_{Z}=91.18 \mbox{ GeV}.
\label{alphas:quarkonia}
\ea
The value of $(\alpha_s)_\tau$ is determined from measurements
of $\tau$ decays~\cite{Baikov:2008jh}; $(\alpha_s)_{Q\bar{Q}}$ comes from heavy quarkonium 
decays~\cite{Amsler:2008zzb}; and $(\alpha_s)_{\rm lattice}$ 
is obtained from lattice computations \cite{Amsler:2008zzb}. 

The $\tau$ decay and heavy-quarkonium determinations  of $\alpha_s$ can be
reasonably assumed to be independent of gluino effects.   Even if very light
gluinos ($\approx 10$ GeV) were present, the value of $\alpha_s$ in
these measurements would not be affected. The lattice QCD value
$(\alpha_s)_{\rm lattice}$ is also determined at $Q<10$ GeV 
from the energy levels of heavy
quarkonia~\cite{Amsler:2008zzb}, and then 
evolved by the authors to $Q=M_Z$ assuming
the SM $\beta$-function. We reconstruct the ``directly measured''
lattice QCD value at $Q=5$ GeV (independent of the gluino effects) 
by backward SM evolution. We then combine the lattice QCD value 
with the other two low-$Q$ measurements, evolved to the same scale
using the SM $\beta$-function, to obtain a composite data input to the fit.

\subsection{$Z$ pole constraints \label{sec:Zpole}}
If $m_{\tilde g} < M_Z$, the value of $\alpha_s(M_Z)$ extracted 
from the LEP $e^+e^-$ hadroproduction data could differ 
from the value obtained from SM fits.  
On the other hand, various determinations of $\alpha_s(M_Z)$ from $Z$
boson width and hadronic event 
shapes~\cite{:2003ih,Becher:2008cf,Davison:2008vx, Dissertori:2009ik, Abbate:2010vw, Abbate:2010xh}
show no obvious need for  BSM 
contributions.  Thus, if gluinos are lighter than $Z$ bosons, their
contributions to the LEP observables are of the order 
of theoretical uncertainties from other sources. 
Notably related to assumptions about nonperturbative hadronization in 
LEP observables, these uncertainties 
remain substantial and produce central values of $\alpha_s(M_Z)$ 
ranging from 0.1135 \cite{Abbate:2010vw, Abbate:2010xh} to 0.1224 \cite{Dissertori:2009ik}.
To deal with this issue of choice, one solution is  
to include available values of $\alpha_s(M_Z)$ 
derived from the $Z$ width and/or 
event shape measurements, assuming that gluino contributions
for these measurements are comparable with the current experimental
plus theoretical uncertainties.

Absent a gluino lighter than the $Z$ boson (i.e., if only SM particles
contribute at $Q<M_Z$), NLO evolution of the composite low-$Q$ value 
in Eq.~(\ref{alphas:lowQ}) to the $Z$ pole results in $\alpha_s(M_Z)$ close 
to 0.118.   Global analysis of hadronic scattering alone also leads to a preferred 
value $\alpha_s(M_Z) = 0.118 \pm 0.005$ at 90\% C.L., 
cf. recent CTEQ fits~\cite{Lai:2010nw}.

\begin{figure}[t]
\includegraphics[width=8cm, angle=-90]{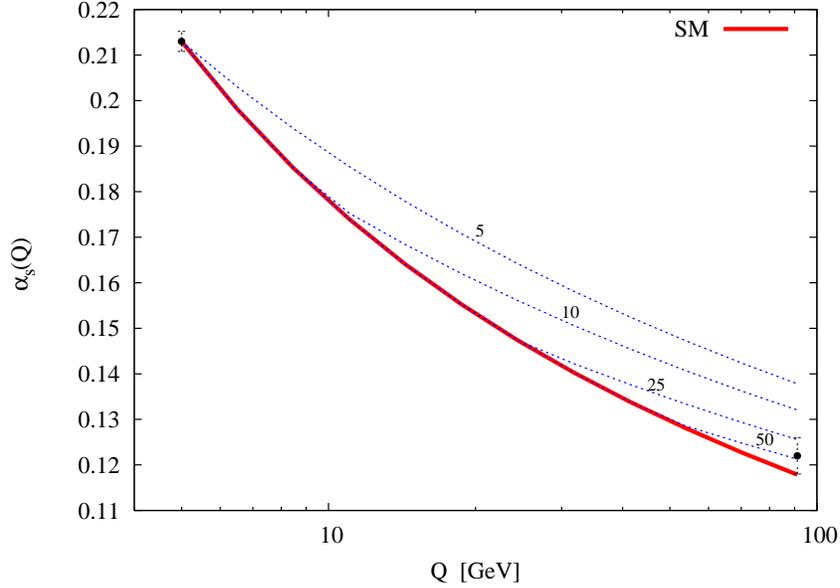}
\caption{\label{fig:run_gino} Running of the strong coupling as a function of the scale $Q$.
The red solid line represents the SM evolution, while the dashed lines are
plotted for $m_{\tilde g}=50,25,10,5$ GeV. The points with the error bands
represent the low-$Q$ and high-$Q$ constraints, given in
Eqs.~(\ref{alphas:lowQ}) and (\ref{alphas:highQ}), respectively.}
\end{figure}

If the gluino is lighter than $M_Z$, the resulting evolved
value at $Q=M_Z$ is higher.  For example, the evolved $\alpha_s(M_Z)$ is
0.126 or 0.121, if $m_{\tilde g}$ is  20 or 50 GeV.  This variation is illustrated in 
Fig.~\ref{fig:run_gino}, showing the dependence of $\alpha_s(Q)$ on the
scale $Q$ in the absence of light gluinos (solid line) and with
gluinos of mass $m_{\tilde g}=50,25,10$ and $5$~GeV.  In the figure, we show
the low-$Q$ constraint (the left data point), as well as one of
available constraints at the $Z$ pole, $\alpha_s(M_Z) =0.123\pm 0.004$
\cite{Dissertori:2009ik}. As seen in the figure, a light gluino with a mass of
$m_{\tilde g}=10$ GeV cannot simultaneously accommodate the low-$Q$
and high-$Q$ constraints. On the other hand, gluinos with mass 
about 50 GeV are compatible with both constraints, and are even
preferred if the high-$Q$ constraint on $\alpha_s$ is larger than 0.118. 

To illustrate typical possibilities, we therefore present two kinds of fits in this 
paper:  one in
which a fixed value of $\alpha_s(M_Z)=0.118$ is assumed; 
and the other in which $\alpha_s(M_Z)$ varies and is constrained by
an assumed high-$Q$ data point, 
\be 
\alpha_s(M_Z) =0.123\pm 0.004, \label{alphas:highQ}
\ee 
compatible with \cite{Dissertori:2009ik}.

\subsection{Log-likelihood function for coupling strength constraints \label{sec:chi2}}

With the additional constraints on the running
coupling, the total log-likelihood function $\chi^2_{tot}$ is 
\ba
\chi^2_{tot}=\chi^2_{h.s.} + \chi^2_{\alpha_s}, \label{chi2tot}
\ea
where $\chi^2_{h.s.}$ is the $\chi^2$ contribution of the hadron
scattering (h.s.) experiments, i.e., DIS, vector boson production, and
jet production; and  $\chi^2_{\alpha_s}$ is the contribution
from the direct constraints on $\alpha_s$: 
\be
\chi^2_{\alpha_s}= \lambda  \sum_{i=1}^{N_{\alpha_s}} 
\left(\frac{\alpha_s^{(i)}\vert_{exp}  -
\alpha_s^{(i)}\vert_{th}}{\delta \alpha_s^{(i)}\vert_{exp}}\right)^2
. \label{chi2alphas}
\ee
In this equation, $N_{\alpha_s}$ is the number of data points constraining $\alpha_s$;
$N_{\alpha_s}=2$ in our case. $\alpha_s^{(i)}\vert_{exp}$ and $\delta \alpha_s^{(i)}\vert_{exp}$
are the central value and error of the 
experimental measurements in Eqs.~(\ref{alphas:lowQ}) 
and (\ref{alphas:highQ});
$\alpha_s^{(i)}\vert_{th}$ are the respective two-loop theoretical values.
We assume that an 
increase in $\chi^2$ by 100 units above the best fit value corresponds 
to approximately 90\% C.L. error, in accordance
with the convention of the previous CTEQ6 analysis
\cite{Pumplin:2002vw} and 2004 gluino study \cite{Berger:2004mj}.   
To match this convention, the $\alpha_s$ contribution $\chi^2_{\alpha_s}$ is
included with a factor $\lambda=37.7$, so that a deviation 
of $\alpha_s^{(i)}\vert_{th}$ by $1.6 \delta
\alpha_s^{(i)}\vert_{exp}$ (90\% C.L.) corresponds to $\Delta
\chi^2_{\alpha_s} \approx 100$.

\section{Global fits\label{sec:results}}

In this section we describe our simultaneous global fit to hadronic scattering data 
and $\alpha_s(Q)$, and we examine  the effects of an additional gluino degree of freedom 
on the PDFs.  

Our SUSY fits include the same set of data as the latest CT10 fit of
parton distributions~\cite{CT10}.  A total of 2753 data points from 35
experiments is included.
Besides the data studied in the previous CTEQ6.6
analysis~\cite{Nadolsky:2008zw}, the new analysis 
includes the combined DIS data from \hbox{HERA-1}~\cite{2009wt} 
and single-inclusive jet data from the Tevatron Run-2 analyses
\cite{Aaltonen:2008eq,Abulencia:2007ez,2008hua}.
The new data provide important constraints on the gluon PDF, the
parton density that is most affected by the gluinos.
The charm and bottom PDFs are also affected, since they are generated
by DGLAP evolution from the gluon PDF above the
initial scale $Q_0 = m_c = 1.3 $ GeV. 
%

\begin{figure}[t]
\begin{center}
\includegraphics[width=5.5cm, angle=-90]{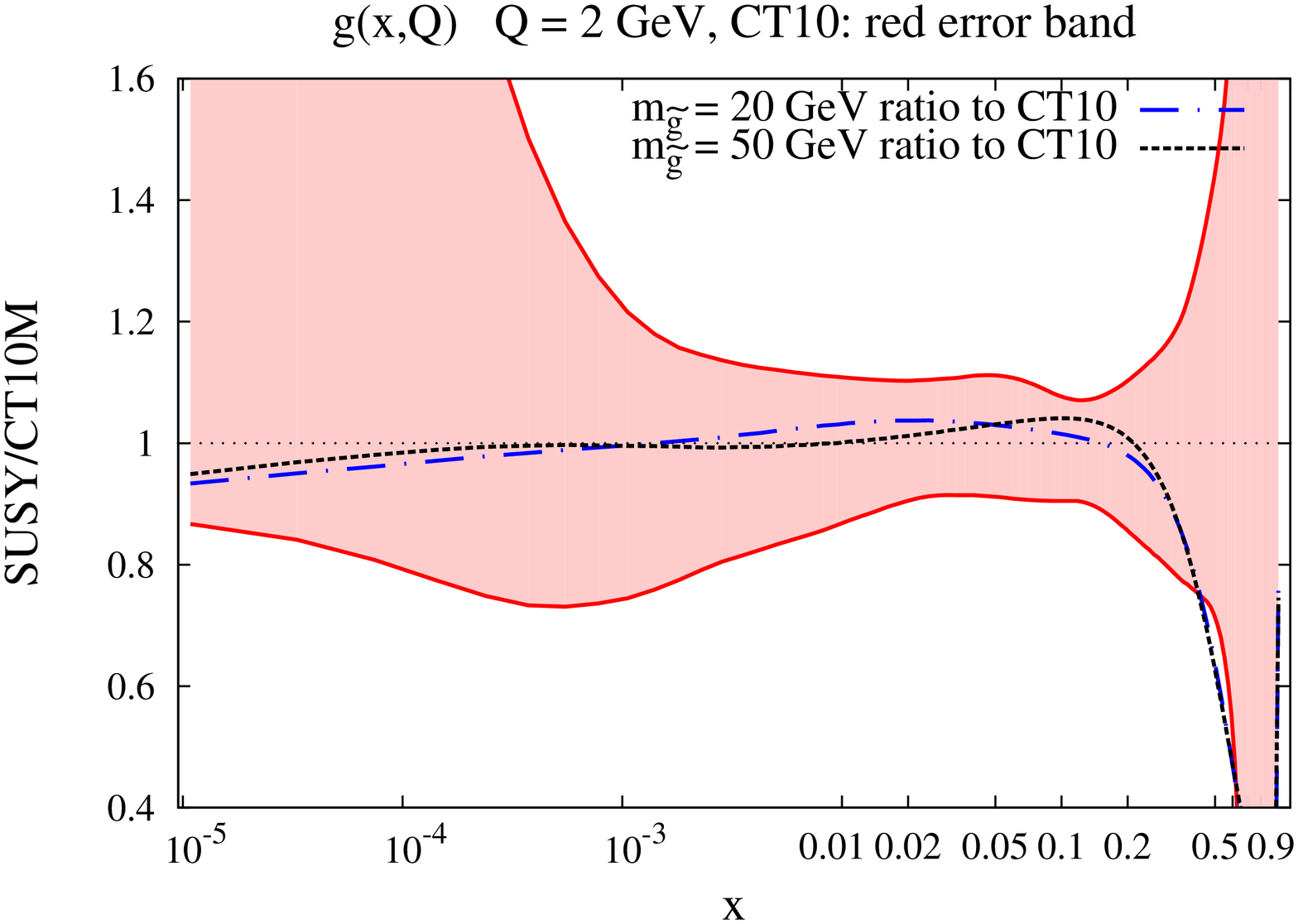}
\includegraphics[width=5.5cm, angle=-90]{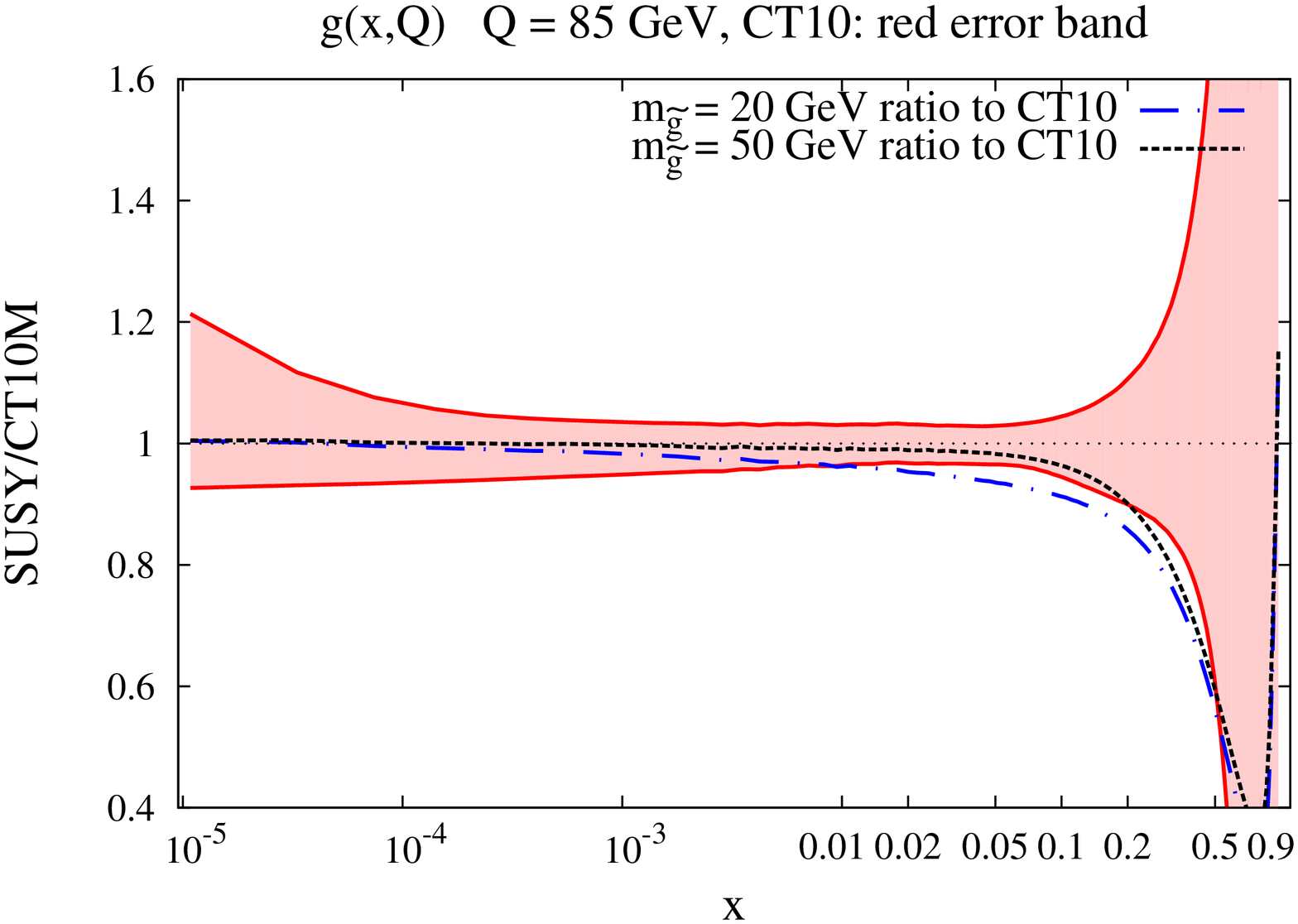}\\
\includegraphics[width=5.5cm, angle=-90]{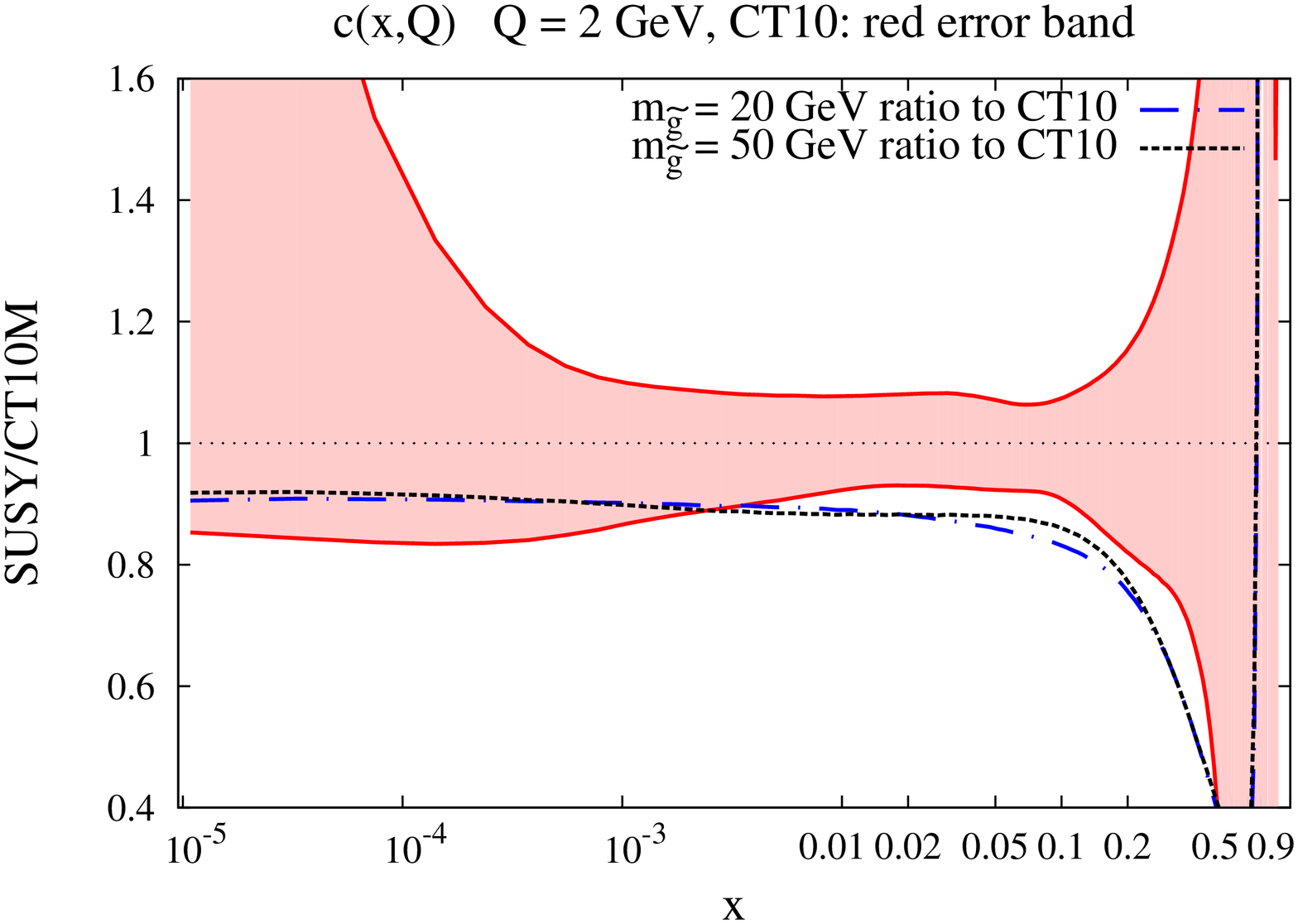}
\includegraphics[width=5.5cm, angle=-90]{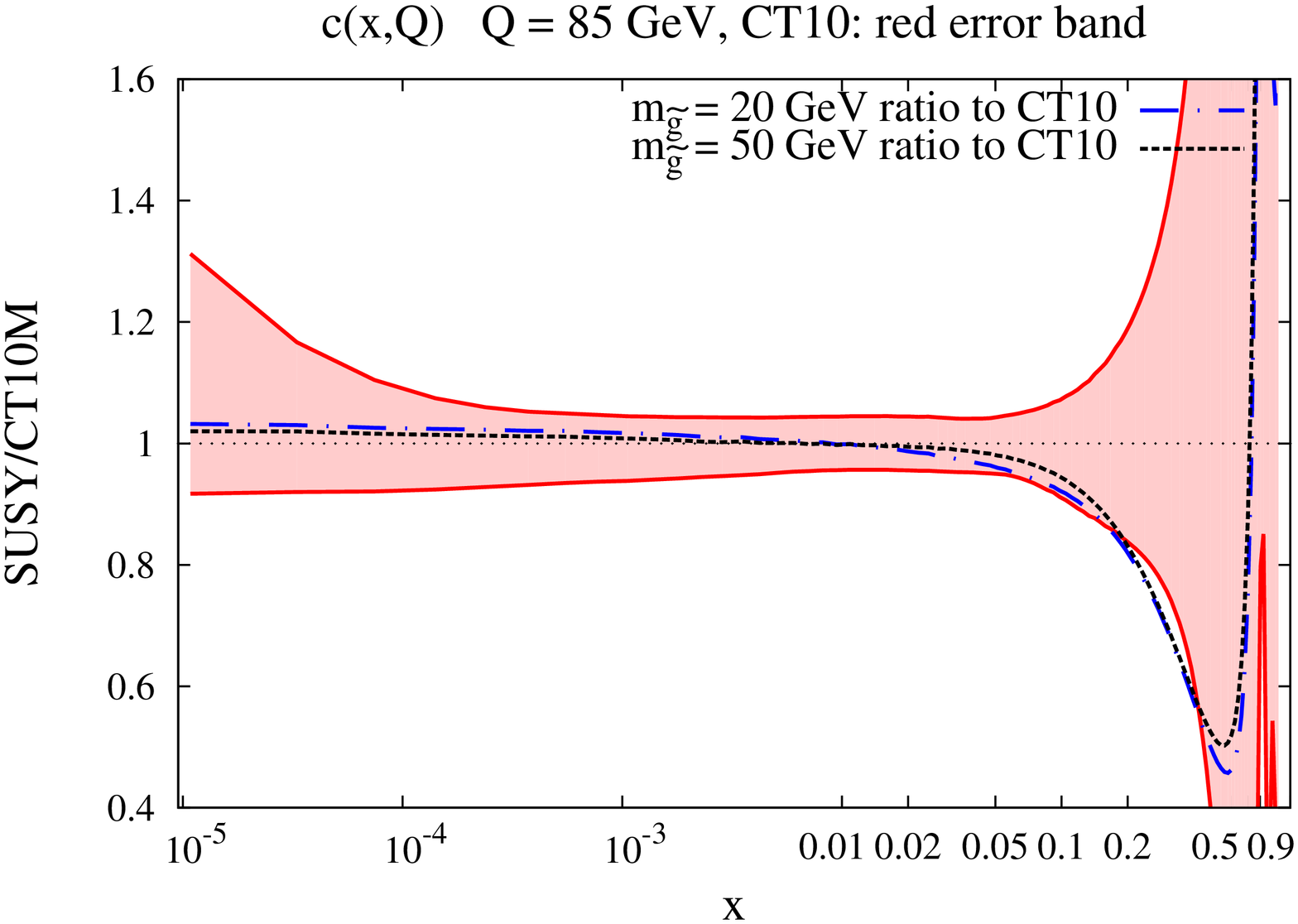}
\caption{\small{Ratios of $g(x,Q)$ (upper row) and $c(x,Q)$ (lower
row) distributions in SUSY fits with floating $\alpha_s(M_Z)$ and the CT10 fit 
at $Q=2$ GeV (left) and $Q=85$ GeV (right), for the gluino mass
$m_{\tilde g}$ of 20 and 50 GeV.}}
\label{fig:PDFs:floating}
\end{center}
\end{figure}

\begin{figure}[t]
\begin{center}
\includegraphics[width=5.5cm, angle=-90]{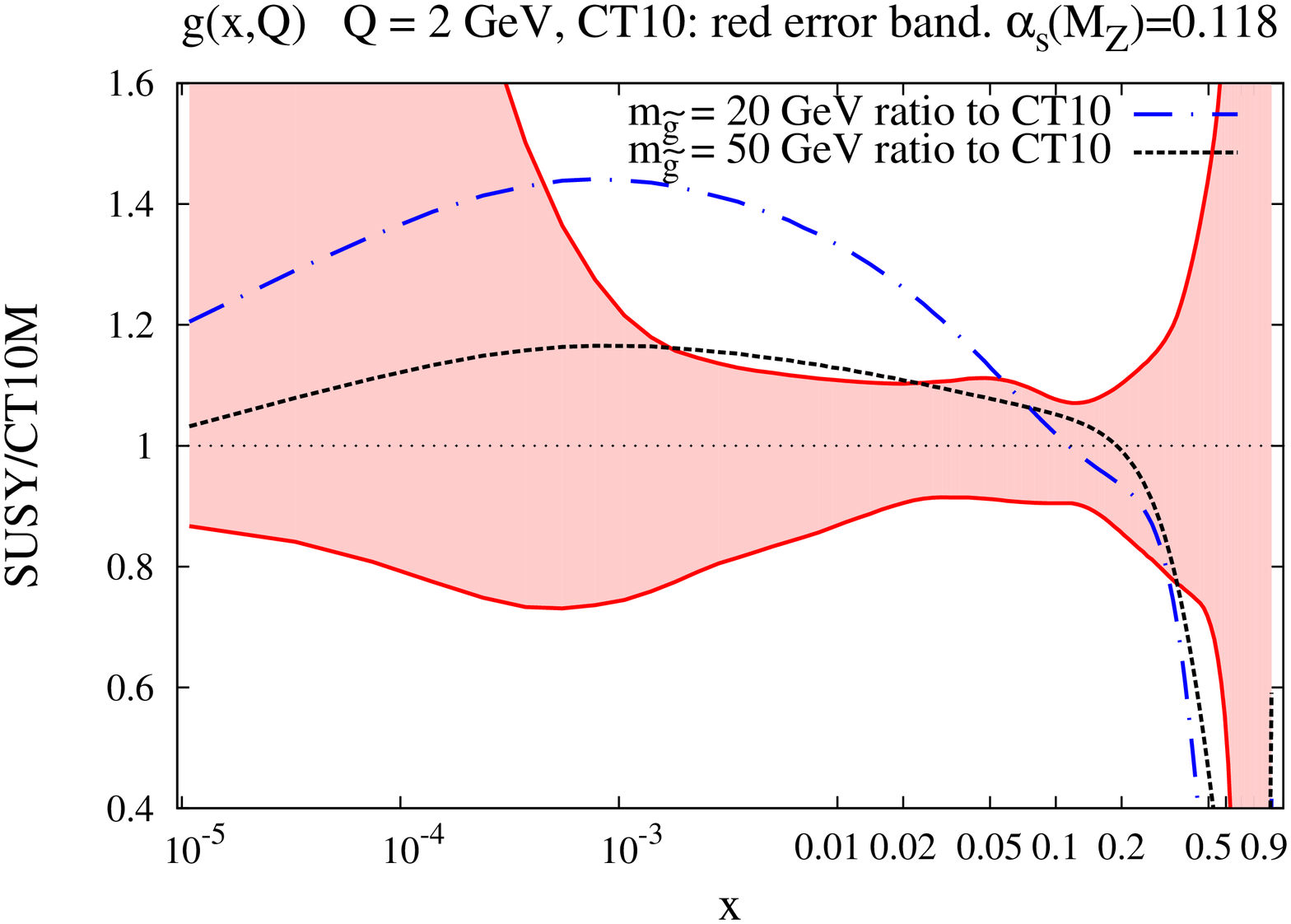}
\includegraphics[width=5.5cm, angle=-90]{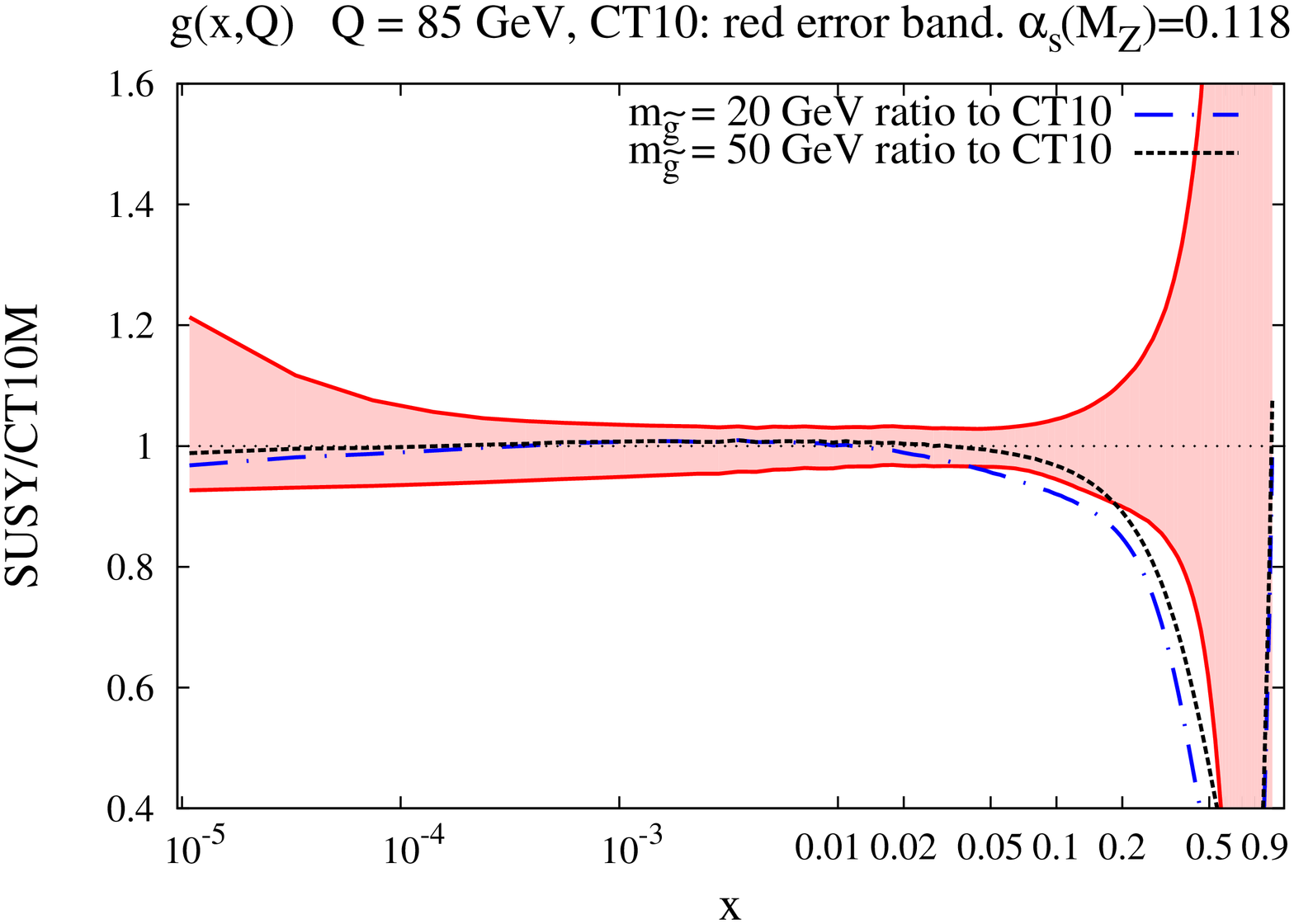}\\
\includegraphics[width=5.5cm, angle=-90]{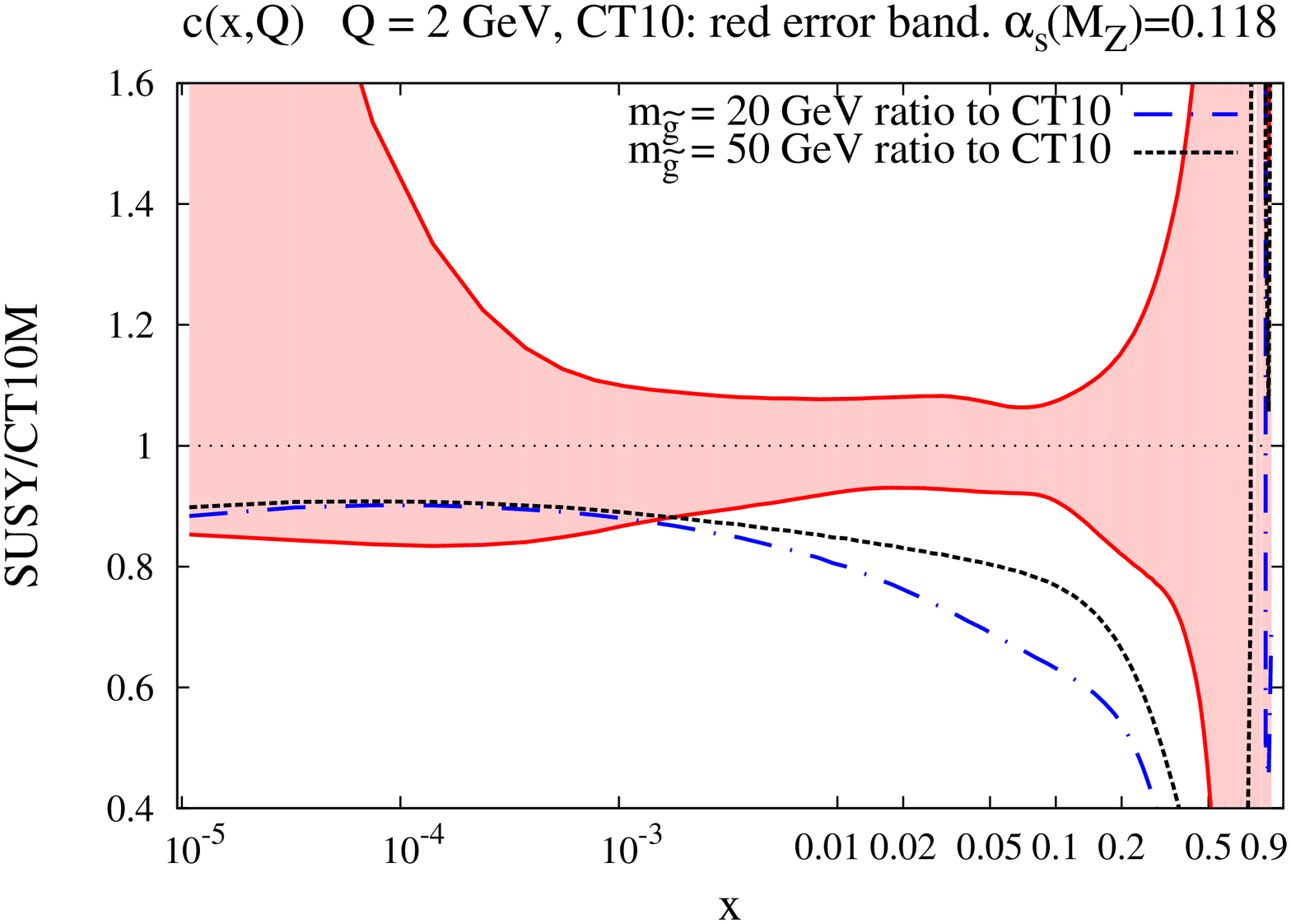}
\includegraphics[width=5.5cm, angle=-90]{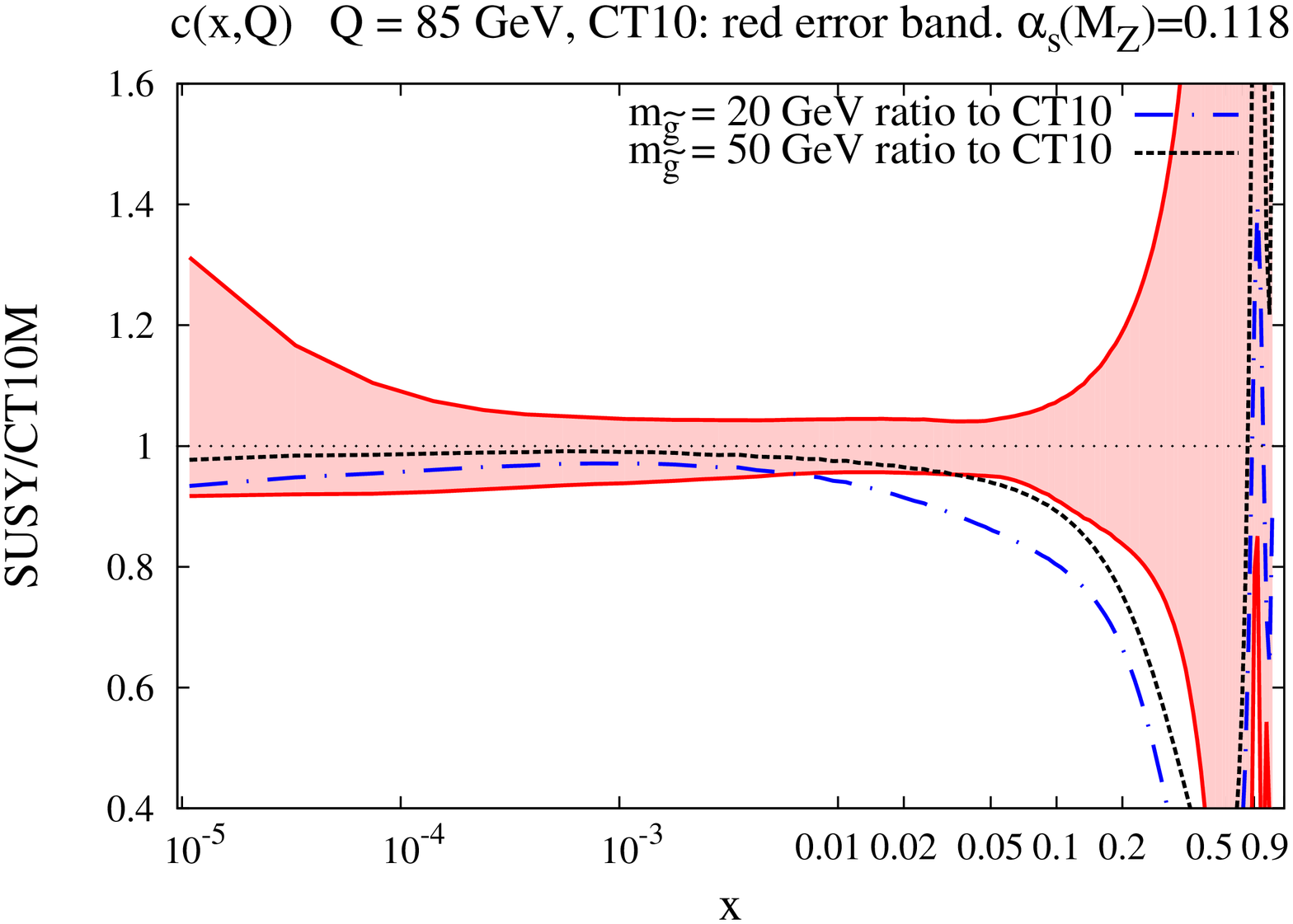}
\caption{\small{Same as Fig.~\protect\ref{fig:PDFs:floating}, but for a
fixed $\alpha_s(M_Z)=0.118$.}}
\label{fig:PDFs:fixed}
\end{center}
\end{figure}

The ratios of the best-fit gluon and charm PDFs in the SUSY sets to their counterparts in 
the standard model CT10 set, $f_{SUSY}(x,Q)/f_{CT10}(x,Q)$,
are shown as dashed curves in
Figs.~\ref{fig:PDFs:floating} and \ref{fig:PDFs:fixed}, at $Q=2$  and 85 GeV, 
for two values of the gluino mass, $m_{\tilde g}=$ 20 and 50 GeV.  
The  normalized CT10 uncertainty bands are shown also, defined as 
\be
\frac{f_{CT10}(x,Q) \pm \delta_{\pm} f_{CT10}(x,Q)}{f_{CT10}(x,Q)} 
\ee
in terms of asymmetric PDF uncertainties 
$\delta_{\pm} f_{CT10}(x,Q)$ \cite{Nadolsky:2001yg,CT10}.
Figure~\ref{fig:PDFs:floating} pertains to fits with a floating
$\alpha_s(M_Z)$, whereas Fig.~\ref{fig:PDFs:fixed} is based on 
a fixed $\alpha_s(M_Z) = 0.118$.

If $\alpha_s(M_Z)$ varies (Fig.~\ref{fig:PDFs:floating}),
modifications in the gluon distribution are moderate at most. 
Some differences with the CT10 predictions are observed at large
$x$, notably in the range $x>0.01$ in $g(x,Q)$ at $Q=85$ GeV and in
$c(x,Q)$ at $Q=2$ GeV. The difference is larger
for a lighter gluino with $m_{\tilde g} = $ 20 GeV. 
Other PDFs exhibit smaller differences, all contained in the standard
model uncertainty band.

For a fixed $\alpha_s(M_Z)=0.118$ (Fig.~\ref{fig:PDFs:fixed}), the differences with CT10 are substantial. At $Q=$2~GeV, the SUSY PDFs lie outside of CT10 error bands for $x$ as
low as $10^{-3}$.  At $Q=$85 GeV, the difference persists at $x>0.01-0.05$. 
Large differences between the SUSY and CT10 PDF's 
in the case of a fixed strong coupling are attributed to sizable
deviations from SM running of $\alpha_s$ and compensating adjustments
in $g(x,Q)$ observed 
for relatively light gluinos; cf. Figs. 12b and 5 
in Ref.~\cite{Berger:2004mj}.

The effect of the gluino on the standard model quark and gluon PDFs can be significant, even if $m_{\tilde g}$ is large compared 
to $m_{c}$ and $m_{b}$.\footnote{We use $m_{c}=1.3$ GeV and $m_{b}=4.5$ GeV. The up,
down and strange quark masses $\{ m_{u},m_{d},m_{s})$ do not play
a role in the evolution, as they are less than the initial evolution
scale $Q_{0}=1.3$ GeV. %
}
Because the gluino is an active constituent of the proton, it carries 
a finite momentum fraction, taken
from the other non-SUSY partons, primarily the gluon. 
This feature is evident in Table~\ref{tab:mom100} where we display the 
partonic momentum
fractions for gluino masses $m_{{\tilde g}}=\{ 20, 50, 100\}$~GeV.

\begin{table}[t]
\begin{center}
\begin{footnotesize}
\begin{tabular}{|c||c|c|c|c|c|c|c|c|c|}
\hline
\multicolumn{10}{|c|}{Momentum fractions for $Q=100$ GeV in percent}
\tabularnewline
\hline
$m_{{\tilde g}} ~[\textrm{GeV}]$&
${\tilde g}$&
$\bar{d}$  &
$\bar{u}$  &
$g$  &
$u$  &
$d$  &
$s$  &
$c$  &
$b$\tabularnewline
\hline
\hline
$20$&
$2.8$&
$3.9$&
$3.4$&
$44.3$&
$21.8$&
$11.4$&
$3.0$&
$1.8$&
$1.2$
\tabularnewline
\hline
$50$&
$1.2$&
$3.9$&
$3.4$&
$45.8$&
$21.8$&
$11.4$&
$3.1$&
$1.9$&
$1.2$
\tabularnewline
\hline
$100$&
$0$&
$3.9$&
$3.4$&
$47.1$&
$21.7$&
$11.4$&
$3.0$&
$1.9$&
$1.2$
\tabularnewline
\hline
\end{tabular}
\end{footnotesize}
\end{center}
\caption{\small Momentum fraction 
$F_{i}=\int_{0}^{1}dx\, x\, f_{i}(x,Q)$ for
each partonic flavor $i$ at scale $Q=100$~GeV. 
Momentum fractions for $\{\bar{s},\bar{c},\bar{b}\}$ 
are not shown and must be included to satisfy the sum rule. \label{tab:mom100}}
\end{table}

Gluinos draw most of their momentum fraction from the gluon, since the primary 
coupling is via the process $g\rightarrow{\tilde g}{\tilde g}$.  
The influence on the quarks is a second-order effect transmitted through
the gluon.  At $Q = 100$~GeV, the momentum fraction of the lighter 
gluinos ($m_{{\tilde g}}\sim 20$~GeV) is comparable to that of the strange quark, 
even though the gluino
mass is an order of magnitude larger.  For $m_{{\tilde g}}\sim 50$~GeV, 
the momentum fraction of the gluino is comparable to that of the bottom 
quark.  The magnified impact of the gluino on the QCD evolution, compared
to the usual quark flavors, can be understood from a comparison of
the $g\rightarrow \tilde g$ splitting kernel,
\begin{equation}
P_{g\to{\tilde g}}(x)=3[(1-x)^{2}+x^{2}],\label{eq:Pgluino}
\end{equation}
 with the usual gluon-quark splitting function \begin{equation}
P_{g\to q}(x)=\frac{1}{2}[(1-x)^{2}+x^{2}].\label{eq:Pquark}
\end{equation}
The effect of the gluino as a hadronic constituent in the QCD evolution
is thus equivalent to that of 6 quark flavors, 
$P_{g\to{\tilde g}}=6\, P_{g\to q}$.

%
\begin{figure}[t]
\begin{center}
\includegraphics[width=8cm,angle=-90]{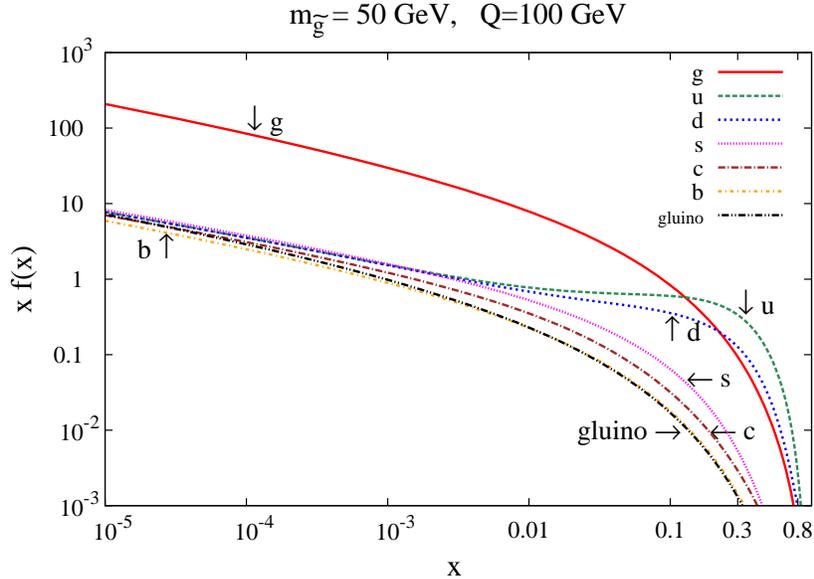} 
\caption{\small{PDFs for various flavors at $Q$=100~GeV, 
for $m_{{\tilde g}}=50$~GeV.} 
\label{fig:gluinoPDF} }
\end{center}
\end{figure}

As an illustration of the relative magnitude of the gluino PDF,
 Fig.~\ref{fig:gluinoPDF} displays PDFs for various parton flavors as a function
of $x$ for our $m_{{\tilde g}}=50$~GeV PDF set and 
a hard scale of $Q=100$~GeV. At $x>0.001$, the gluino PDF is about equal 
to the bottom quark PDF, the smallest of the quark PDFs.  For smaller $x$, 
it grows in magnitude and catches up with the other quark PDFs at
 $x=10^{-5}$, as a consequence of its faster DGLAP evolution.
Parton-parton luminosities dependent on the gluino PDF, useful for computations of 
cross sections, are plotted in Appendix~\ref{app:lumi}.

\section{Comparison of theory and data\label{sec:compwdata}}

In this section, we show the results of our global fits, the constraints we obtain on the mass 
of a gluino, and the impact of a gluino degree of freedom on the analysis of jet data. 

The figures in the previous section show that the SM+SUSY PDFs disagree with CT10 
PDFs if gluinos are lighter than 20 GeV, indicating that 
the SM+SUSY PDFs for these gluino masses cannot describe the global hadronic data 
well.   Gluinos with somewhat larger masses can be
accommodated, or may be slightly preferred to the pure SM case,
depending on the value of $\alpha_s(M_Z)$.  These points are illustrated in 
a different way by the summary of values of $\chi^2$ in 
Table~\ref{table:FitResults}, for $m_{\tilde g}=$10, 20, and 50 GeV,  
as well as for the standard model case (equivalent to $m_{\tilde g}=\infty$).  
\begin{table}[t]
\begin{center}
\begin{footnotesize}
\begin{tabular}{|c||c|c|c|c|c|}
\hline
\multicolumn{6}{|c|}{SUSY analysis with a fixed $\alpha_s(M_Z)=0.118$}
\tabularnewline
\hline
$m_{{\tilde g}} ~[\textrm{GeV}]$&
$\chi^2_{h.s.}$&
$\chi^2_{tot}$  &
$\chi^2/\textrm{npt}$: HERA-1    &
$\chi^2/\textrm{npt}$: jet prod. &
$\alpha_s(M_Z)$ 
\tabularnewline
\hline
\hline
$10$&
$3154$&
$12550$&
$1.31$&
$1.24$&
$0.118$
\tabularnewline
\hline
$20$&
$3030$&
$7882$&
$1.24$&
$1.19$&
$0.118$
\tabularnewline
\hline
$50$&
$2923$&
$3788$&
$1.18$&
$1.10$&
$0.118$
\tabularnewline
\hline
$\infty$&
$2918$&
$3004$&
$1.16$&
$1.09$&
$0.118$
\tabularnewline
\hline
\multicolumn{6}{|c|}{SUSY analysis with a floating $\alpha_s(M_Z)$}
\tabularnewline
\hline
$m_{{\tilde g}} ~[\textrm{GeV}]$&
$\chi^2_{h.s.}$&
$\chi^2_{tot}$  &
$\chi^2/\textrm{npt}$: HERA-1  &
$\chi^2/\textrm{npt}$: jet prod.  &
$\alpha_s(M_Z)$ 
\tabularnewline
\hline
\hline
$10$&
$2892$&
$3124$&
$1.14$&
$1.06$&
$0.132$
\tabularnewline
\hline
$20$&
$2897$&
$2958$&
$1.15$&
$1.06$&
$0.127$
\tabularnewline
\hline
$50$&
$2896$&
$2901$&
$1.15$&
$1.03$&
$0.121$
\tabularnewline
\hline
$\infty$&
$2918$&
$2960$&
$1.16$&
$1.09$&
$0.118$
\tabularnewline
\hline

\end{tabular}
\end{footnotesize}
\caption{\small $\chi^2$ values in the global analyses with a floating
  and fixed $\alpha_s(M_Z)$, for various gluino mass values.\label{table:FitResults}}
\end{center}
\end{table}
The table shows the log-likelihood values $\chi^2_{h.s.}$ and $\chi^2_{tot}$, without and with
the imposition of $\alpha_s$ constraints, 
as defined in Eqs.~(\ref{chi2tot}) and (\ref{chi2alphas}); as
well as $\chi^2$ per number of data points for \hbox{HERA-1} DIS~\cite{2009wt} 
and Tevatron Run-1 and Run-2 single-inclusive jet cross sections~\cite{Affolder:2001fa,Abbott:2000kp,Aaltonen:2008eq,Abulencia:2007ez,2008hua}. 
In the fit with a floating $\alpha_s$, the best-fit $\alpha_s(M_Z)$ 
is also shown.  A comparison of the upper and lower halves of the
table shows that the relation between $\chi^2$ and $m_{\tilde g}$
depends on whether $\alpha_s(M_Z)$ is fixed or floating.

{\bf Fixed $\alpha_s(M_Z)$. } 
In a fit with a fixed $\alpha_s(M_Z)$, only constraints 
from the hadronic data, associated with the term $\chi^2_{h.s.}$ 
(and not with the total $\chi^2_{tot}$) play a meaningful role. 
The upper half of Table~\ref{table:FitResults} shows $\chi^2$ values
from a fit with fixed $\alpha_s(M_Z)=0.118$.\footnote{This value, compatible
with the current world average, is about $1\sigma$ below 
$\alpha_s(M_Z)=0.123\pm0.004$; hence, the SM
fit with this $\alpha_s(M_Z)$ value has a higher $\chi^2_{tot}$ (in
the last line of the upper table) than a fit with a floating
$\alpha_s(M_Z)$ (in the last line of the lower table).}
In this case, the gluino's effect of slowing the evolution of $\alpha_s(Q)$ from 
$Q=M_Z$ to $Q=5$~GeV runs into strong disagreement with the low-$Q$ constraint;
$\chi^2_{tot}$ grows quickly as $m_{\tilde g}$ decreases, corresponding to a
difference of many standard deviations between the measured and
predicted $\alpha_s$ values at $Q=5$ GeV. More importantly, the
hadronic data by themselves disfavor very light gluinos, with
$m_{\tilde g}=25$~GeV or less excluded according to the criterion  
$\Delta \chi^2 \equiv 
\chi^2_{SUSY}(m_{\tilde g}) - \chi^2_{CT10}
< 100$ applied to $\chi^2_{h.s.}$.

\begin{figure}[t]
\begin{center}
\includegraphics[width=9cm,angle=-90]{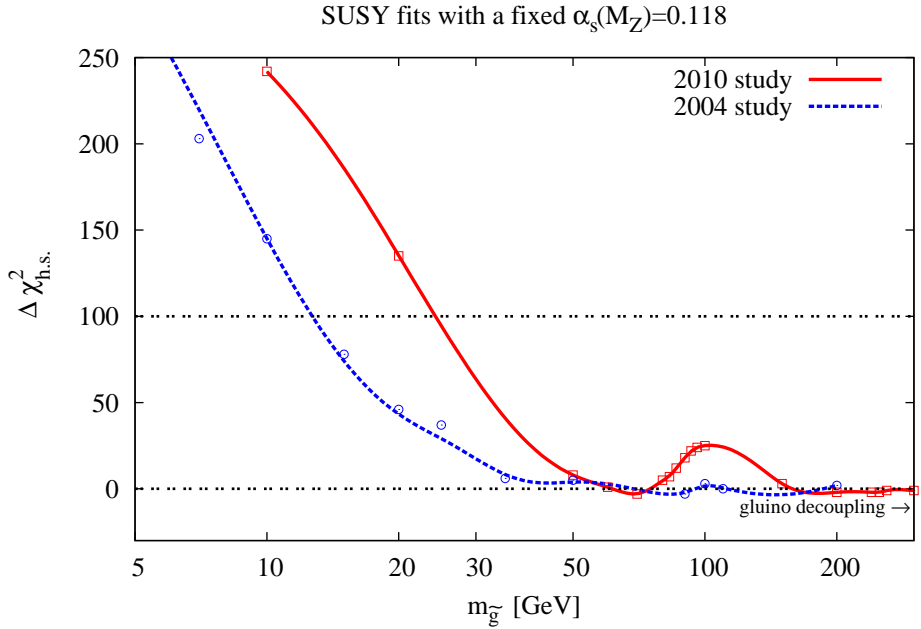}
\caption{\small{Values of $\Delta\chi^2_{h.s.}$ 
vs. $m_{\tilde g}$ 
 are shown for a fixed value of $\alpha_s(M_Z) = 0.118$
for the 2004 study (blue
dashed line) and our new one (red solid line).}}
\label{chisq1}
\end{center}
\end{figure}

\begin{figure}[t]
\begin{center}
\includegraphics[width=9cm,angle=-90]{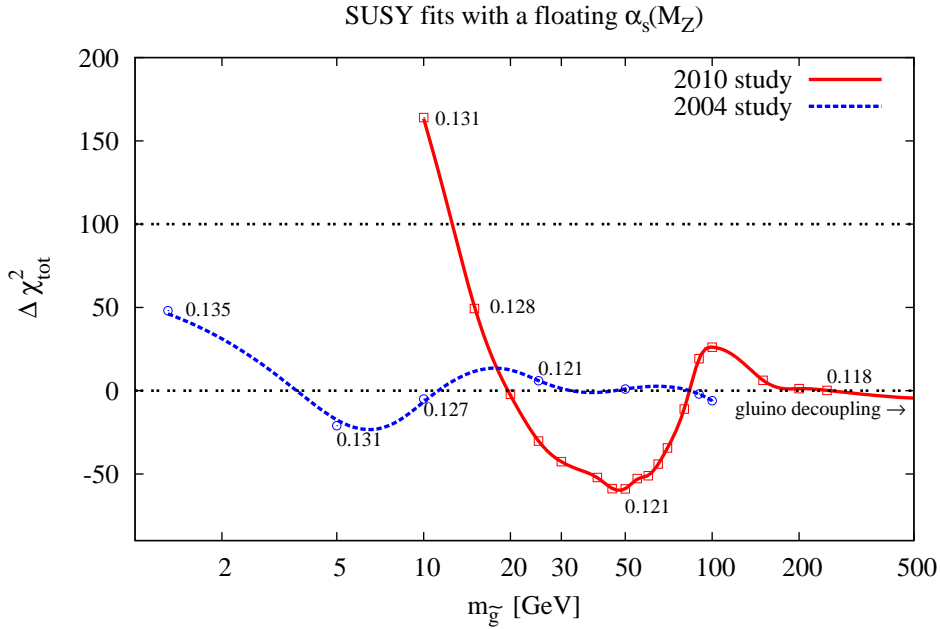}
\caption{\small{Values of $\Delta\chi^2_{tot}$ vs $m_{\tilde g}$ are shown for 
for different values of $\alpha_s (M_Z)$ for the 2004 study (blue
dashed) and our new one (red solid).}}
\label{chisq2}
\end{center}
\end{figure}

\null
\par

{\bf Floating $\alpha_s(M_Z)$.} 
The values in the lower half of Table~\ref{table:FitResults} are for 
SM+SUSY fits with a variable  $\alpha_s(M_Z)$. In this case, 
the constraints from both the hadronic scattering and direct measurements 
of $\alpha_s(M_Z)$ are relevant.  The most meaningful log-likelihood term 
is $\chi^2_{tot} = \chi^2_{h.s.}+\chi^2_{\alpha_s}$.
 If $\alpha_s(M_Z)$ varies, the hadronic scattering data on their own,
including the HERA-1 and Tevatron jet data sets,
are compatible with practically any gluino mass.  The contribution to
$\chi^2$ from the hadron scattering experiments, $\chi^2_{h.s.}$,
stays approximately the same as in the SM case, 
or improves slightly, as gluinos with masses of 10, 20, and 50
GeV are introduced. 

This agreement with the hadronic
scattering data, hardly affected by the gluino mass, results from 
compensation between modifications in the shape of the gluon PDF
and an increase in the preferred $\alpha_s(M_Z)$, which grows from
0.118 in the SM case to 0.132 for $m_{\tilde g}=10$ GeV.  In contrast,
the total likelihood function for the fit to the hadronic scattering data   
and $\alpha_s$ values, introduced as $\chi^2_{tot}$ in
Eq.~(\ref{chi2tot}), 
varies considerably as a function of the gluino mass.
Our assumed high-$Q$ constraint of $\alpha_s(M_Z)=0.123\pm 0.004$ is 
slightly higher than $\alpha_s(M_Z)=0.118$ obtained
by the SM evolution from 
$\alpha_s(Q=5\mbox{ GeV})=0.213 \pm 0.002$ in Eq.~(\ref{alphas:lowQ}). 
This enhanced value of  $\alpha_s(M_Z)$ would favor a slower QCD evolution 
above the gluino mass threshold at about 50 GeV, 
cf.~Fig.~\ref{fig:run_gino}.   Consequently, $\chi^2_{tot}$ is smaller at  
$m_{\tilde g} = 50$ GeV than in the SM case, with the difference
dependent on the choice of the high-$Q$  value of  $\alpha_s(M_Z)$.
 Specifically, we observe that 
$\chi^2_{SUSY}(m_{\tilde g}) - \chi^2_{\rm CT10}$ can be as small as 
$-50$, if we take  
$\alpha_s(M_Z)=0.123\pm 0.004$, but this difference decreases if a
smaller $\alpha_s(M_Z)$ is used for the high-$Q$ constraint.  For lower
gluino masses of 10 or 20 GeV, $\alpha_s(M_Z)$ increases
and  eventually is incompatible with the direct constraints. 

\subsection{$\Delta\chi^2$ as a function of gluino mass} 

The behavior of  $\Delta \chi^2$ in the whole range of gluino masses
is illustrated by Fig.~\ref{chisq1} for a fixed $\alpha_s(M_Z)=0.118$,
and by Fig.~\ref{chisq2} for a floating $\alpha_s(M_Z)$.   The 
quantitative likelihood of a given mass $m_{\tilde g}$ is 
specified by $\Delta\chi^2=\chi^2(m_{\tilde g}) - \chi^2_{\rm CT10}$, the 
difference from the $\chi^2$ value obtained in the CT10 SM fit.  Values 
of $\Delta\chi^2$ in excess of 100 units disfavor an assumed $m_{\tilde g}$ 
at about 90\% C.L. Ref.~\cite{Berger:2004mj,Pumplin:2002vw}, while a 
negative $\Delta\chi^2$ indicates a 
preference for this $m_{\tilde g}$.  Variations in $\Delta\chi^2$ with a 
magnitude below 100 units can result from a variety of sources and are 
generally viewed as not significant enough to warrant strong conclusions.  

In Fig.~\ref{chisq1}, two curves are shown for $\Delta \chi^2_{h.s.}$, 
the difference between the log-likelihoods in the fits 
performed in the SM+SUSY and SM scenarios for $\alpha_s(M_Z)=0.118$.  
Here $\Delta\chi^2_{h.s.}$ is computed  from the hadronic scattering
contribution only, $\chi^2_{h.s.}$.  
The blue (dashed) curve  represents the 2004 analysis~\cite{Berger:2004mj}. 
The red (solid) curve is obtained in the present study, resulting in  
a tighter lower bound on $m_{\tilde g}$.  
The left branch of the 2010 curve intercepts the $\Delta\chi^2_{h.s.}=100$ line at 
$m_{{\tilde g}}\approx 25$ GeV.  The 2004 curve allows for 15 GeV 
gluinos and has a broader valley with respect to the 2010 one.
This figure shows the improvements in the constraints from the
present study, reflecting the inclusion of the latest precise
data and technical advances  in the the CTEQ analysis since the 2004 
publication, including treatment of correlated systematic uncertainties and  
normalization uncertainties.  

Figure~\ref{chisq2} illustrates the fits with a variable $\alpha_s(M_Z)$. 
Two curves are shown for $\Delta \chi^2_{tot}$, 
the difference between the log-likelihoods in the fits 
performed in the SM+SUSY and SM scenarios in 2004 (blue dashed line) 
and 2010 (red solid line). 
Best-fit values of $\alpha_s(M_Z)$ for some gluino masses are indicated
by numerical labels near each curve. 
In this figure, $\Delta\chi^2_{tot}$ is computed from the total function
$\chi^2_{tot}$. It includes the direct
constraints on $\alpha_s(Q)$ in the current study and does not
include the $\alpha_s$ constraint in the 2004 fit. 

The figure emphasizes 
our earlier observation that the direct $\alpha_s$ constraints 
improve the constraining power of the global analysis. 
At $m_{\tilde g}\rightarrow \infty$, 
the fit converges to the pure QCD value 
and $\alpha_s(M_Z)\approx 0.119$.  According 
to the $\Delta \chi^2 \leq 100$ test, gluinos lighter than 15 GeV are 
disfavored for all $\alpha_s(M_Z)$. Gluinos in the mass range 15 to 50~GeV are 
allowed if $\alpha_s(M_Z)$ takes a value in the 
range 0.121 to 0.131. Gluinos heavier than 50 GeV are allowed for practically 
any $\alpha_s(M_Z)$ value. 
By contrast, the 2004 curve exhibits only a  shallow minimum around 5 to 6~GeV, 
and it is relatively flat as compared to the 2010 curve.  The 2004 curve does not 
establish pronounced lower bounds on $m_{\tilde g}$, for a free $\alpha_s(M_Z)$.

The 2010 curve in Fig.~\ref{chisq2} exhibits an intriguing minimum for a gluino of about 
50 GeV, corresponding to $\alpha_s(M_Z)$ of 0.121.   Other that noting it,  we choose 
not to base conclusions on this minimum for two reasons.  First, from the point 
of view of the fit itself, given its initial inputs, we adhere to statement that only values 
of $|\Delta\chi^2|$ in excess of 100 units are considered significant.  Second, the 
depth of this minimum is a reflection of the value of the input constraint 
$\alpha_s(M_Z)=0.123\pm 0.004$.   The dip grows deeper (becomes more shallow) 
if a larger (smaller) value of the direct constraint is taken at $M_Z$. 
For example, gluinos with mass 50 GeV would be disfavored if the direct constraint 
$\alpha_s(M_Z) < 0.118$ were taken, compatible with some existing analyses of LEP 
data in pure QCD~\cite{Abbate:2010vw, Abbate:2010xh}.   Stronger conclusions on $m_{\tilde g}$ 
await an independent reduction in the uncertainties on $\alpha_s(M_Z)$.  

\begin{figure}[tbp]
\begin{center}
\includegraphics[width=10cm, angle=-90]{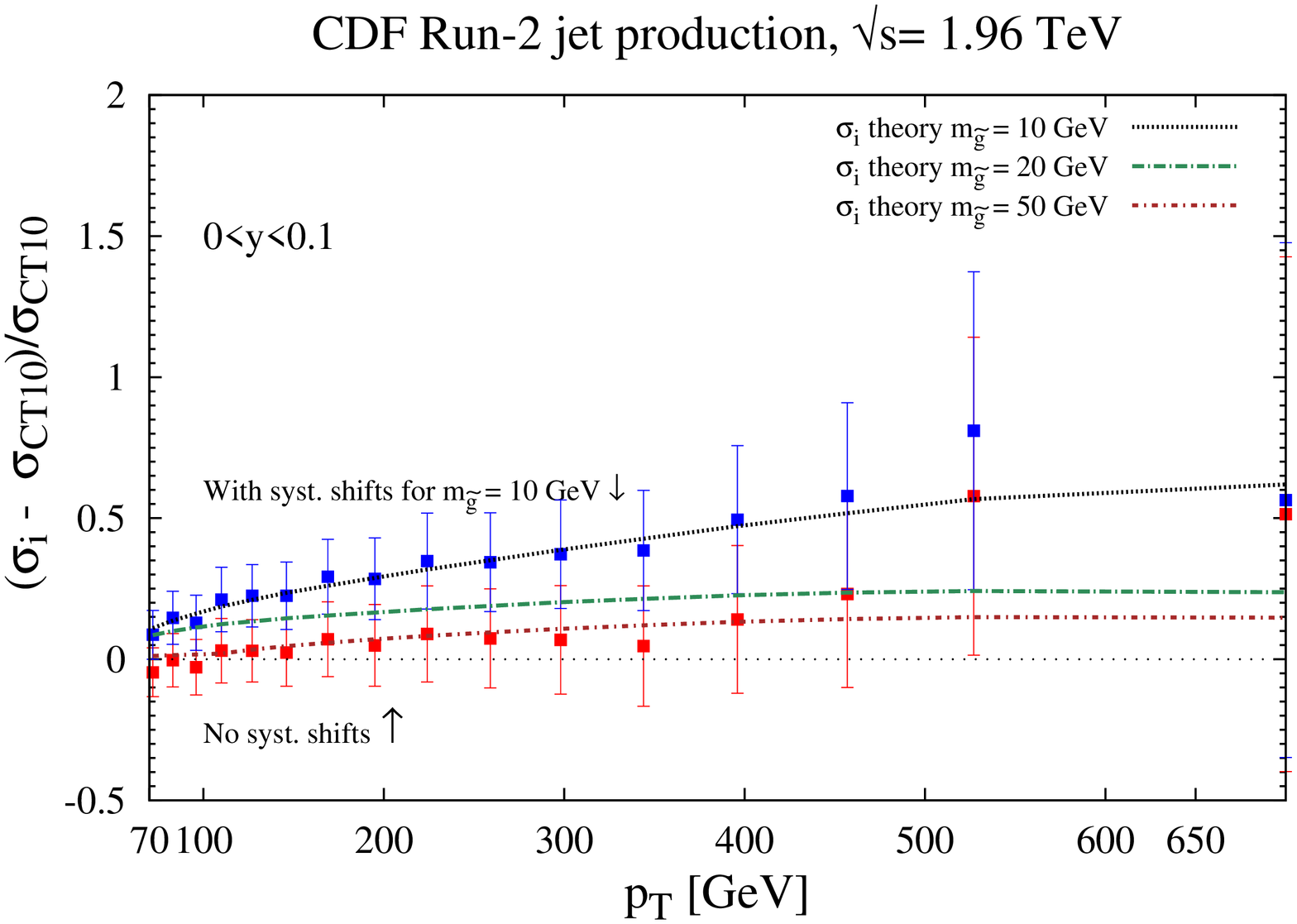}\\
\includegraphics[width=10cm, angle=-90]{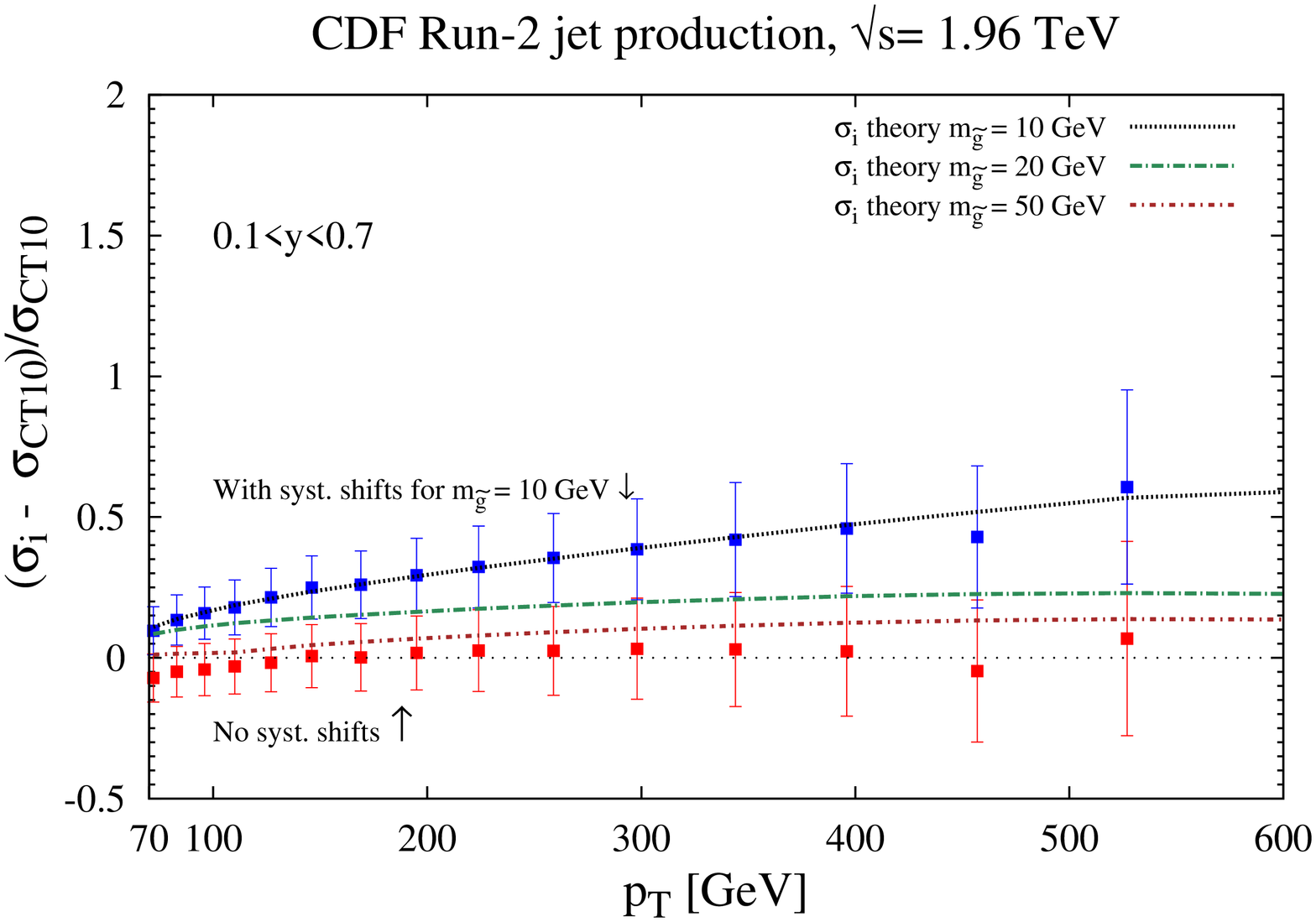}
\caption{Comparison of theoretical predictions for single-inclusive
jet cross sections with experimental data from CDF Run-2 for 
two bins in jet rapidity $y$.}
\label{fig:PtjetCDF}
\end{center}
\end{figure}

\begin{figure}[tbp]
\begin{center}
\includegraphics[width=10cm, angle=-90]{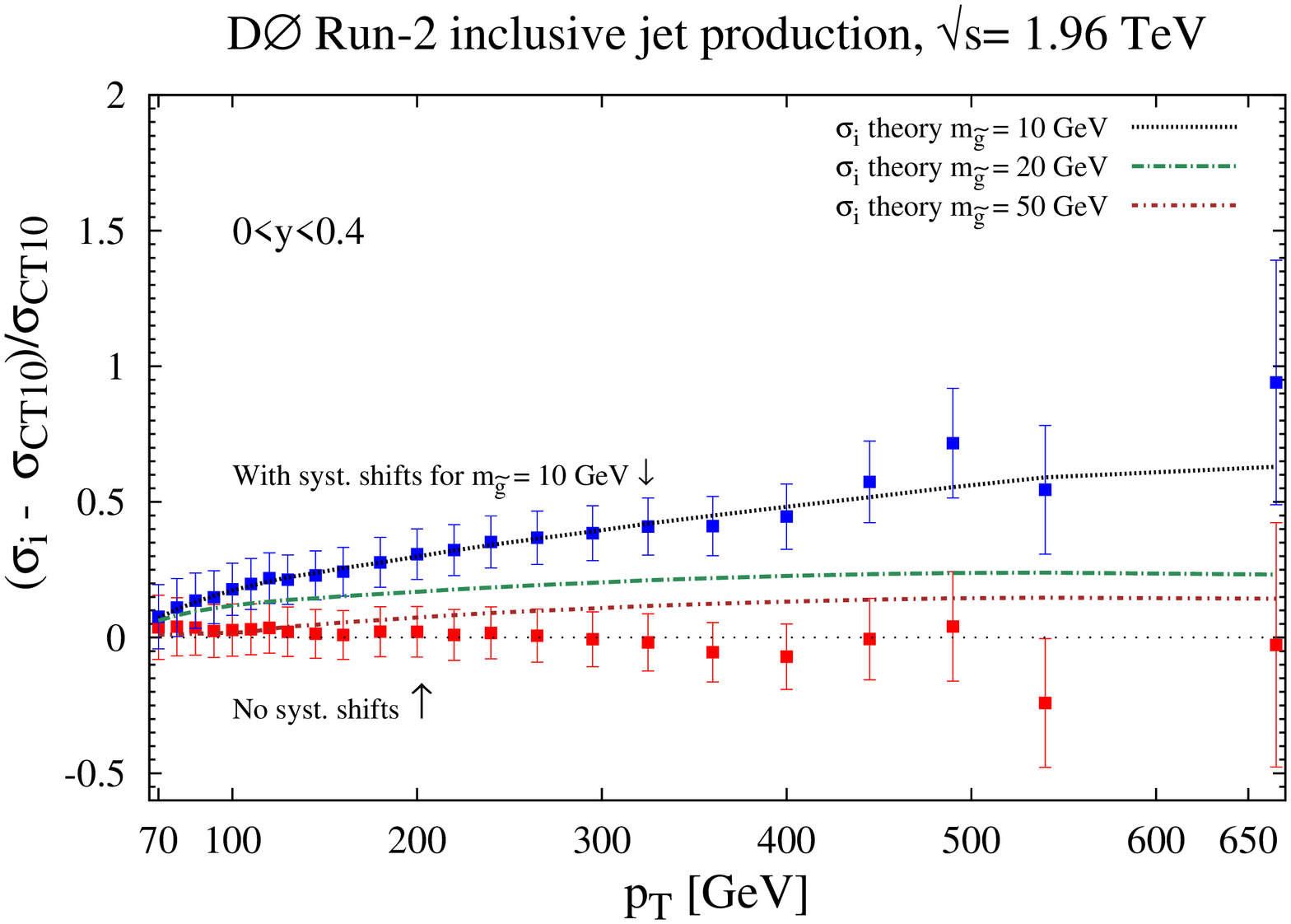}\\
\includegraphics[width=10cm, angle=-90]{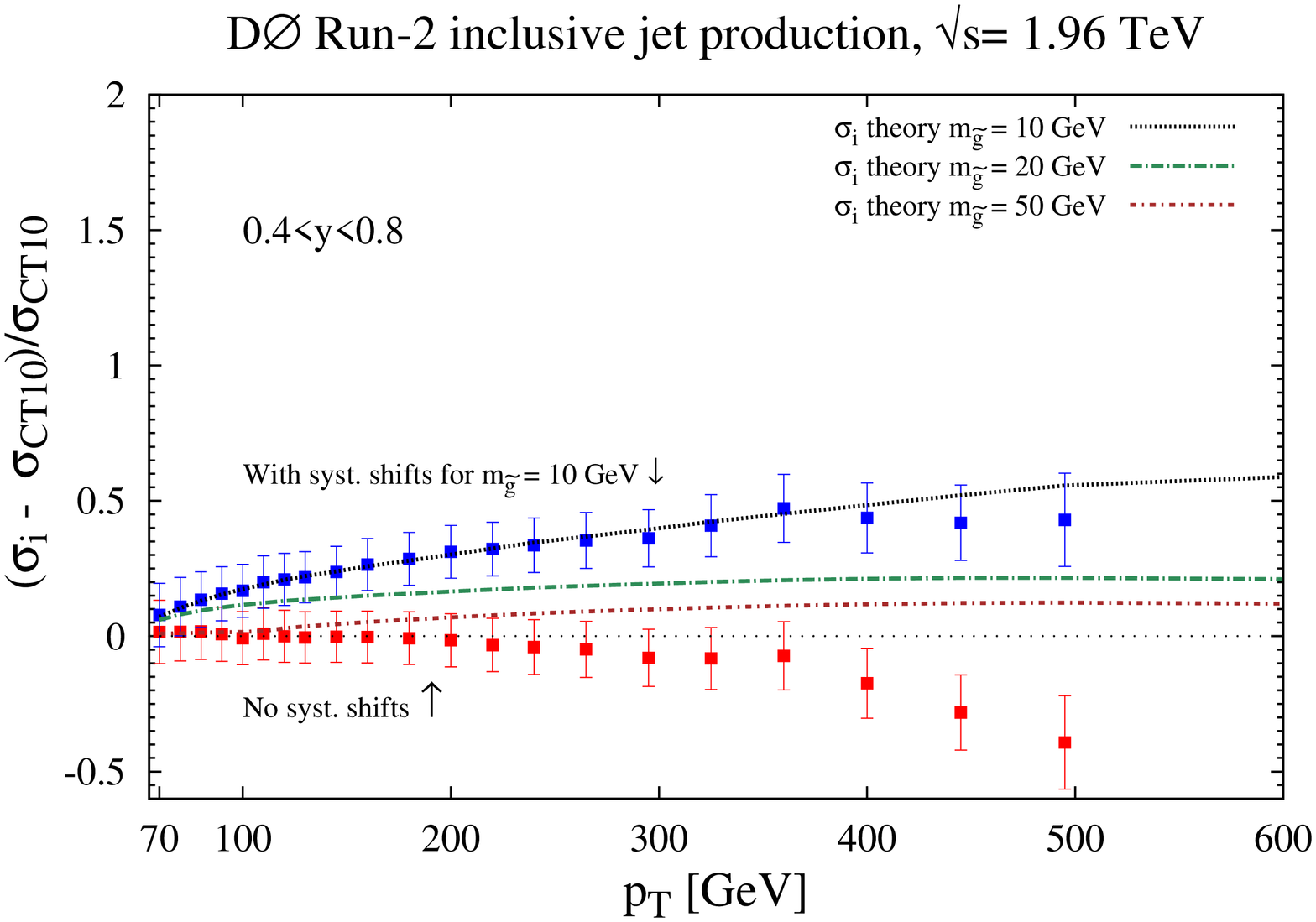}
\caption{Same as Fig.~\ref{fig:PtjetCDF}, for 
two bins in jet rapidity $y$ in the D0 Run-2 jet cross section measurement.}
\label{fig:PtjetD0}
\end{center}
\end{figure}

\subsection{Comparison with Tevatron jet cross sections \label{sec:TevatronJets}}

Table~\ref{table:FitResults} indicates that the hadronic scattering
data, including the combined \hbox{HERA-1} and Tevatron jet
cross sections, may still allow contributions from fairly light gluinos. 
This observation is somewhat counterintuitive with regard to the precise Tevatron
jet cross sections, which could be expected to be sensitive to non-SM
contributions in the strong interaction sector. 
SUSY degrees of freedom introduce new subprocesses in the jet cross
sections, such as $gg\to{\tilde g}{\tilde g}$ 
and $q\bar{q}\to {\tilde g} {\tilde g}$. 
The change in $\alpha_s(Q)$ and the alteration of the gluon and quark PDFs 
also influence the jet rate.   However, jet cross section measurements
are affected by systematic effects that dominate over statistical
uncertainties, notably by the uncertainty on jet energy scale
and jet energy resolution.  Correlated systematic 
shifts in the Tevatron jet data must be taken into account when 
comparisons are made to theory predictions \cite{Pumplin:2009nk}. 
In our study, systematic uncertainties in the jet data limit 
the strength of our conclusions. 

These observations are illustrated by plots of  the CDF Run-2 and 
D$0$ Run-2 data vs. theory in Figs.~\ref{fig:PtjetCDF} and
\ref{fig:PtjetD0}.   Our results are computed with a floating $\alpha_s$.
As reference values, we use SM cross sections computed with the CT10 PDFs.  
Differences from the SM cross section are presented as 
\be
(\sigma_i - \sigma_{\textrm{CT10}})/\sigma_{\textrm{CT10}}, 
\ee
where $\sigma_i$ are the SM+SUSY cross sections computed for gluino
masses of 10, 20, and 50 GeV. The values of the jet  
transverse momentum $p_T$ are displayed along the horizontal axis.  Two 
bins in the rapidity variable $y$ are shown for each experiment; 
the behavior in the rest of the bins is similar.    

The lower (red) error bars represent the unshifted data. The upper (blue)
error bars show the data that are shifted by their systematic
uncertainty so as to maximize the agreement with theory for $m_{\tilde g} =
10$ GeV.  Without the correlated shifts, the data would disfavor the light
gluinos with a mass of 10 GeV. The perspective changes
significantly if systematic shifts are allowed: the line representing
$m_{\tilde g}=$ 10 GeV now lies completely inside the error bars. Similarly,
if $m_{\tilde g}$ is equal to 20 or 50 GeV, the effective shifts of the
data change to achieve acceptable agreement with the theory curve for this
mass.\footnote{The extent of
plausible systematic shifts is determined by matrices of correlated
systematic errors, provided by both Tevatron collaborations and
implemented in the CT09 ~\cite{Pumplin:2009nk} and CT10 
analyses \cite{Lai:2010nw}.}   

The systematic uncertainties make it difficult to disfavor the light gluinos 
solely on the basis of the Tevatron Run-2 jet data.  The figures 
show that the gluino contributions affect 
the whole $p_T$ range, as a result of the momentum sum
rule and other connections between the PDFs of different flavors and
at different $(x,Q)$ values. Modifications in the jet cross sections
due to ``new physics'' associated with the gluinos cannot be isolated
to a specific $p_T$ interval, contrary to the assumptions made in some
experimental studies of jet cross sections~\cite{Abazov:2009nc}.

\section{Cross sections at the LHC\label{sec:LHC}}

The possible existence of color-octet fermions with masses
in the range 30 to 100 GeV, allowed by hadronic data according
to our analysis, raises the prospects for their detection in the extended 
range of transverse momenta at the LHC.  As explained 
in early sections of this paper, these new fermions modify QCD parameters,
primarily the QCD coupling $\alpha_s(M_Z)$ and the gluon and sea-quark
PDFs.  Precise studies of cross sections at LHC energies 
thus have the potential to reveal differences from pure SM QCD, such as 
the presence of color-octet fermions, provided the LHC measurements are 
supplemented by a robust program to reduce
uncertainties in $\alpha_s$, PDFs, and other SM parameters, which may
otherwise reduce sensitivity of the LHC observables to the gluino contributions.

\begin{figure}[tp]
\begin{center}
\includegraphics[width=11cm]{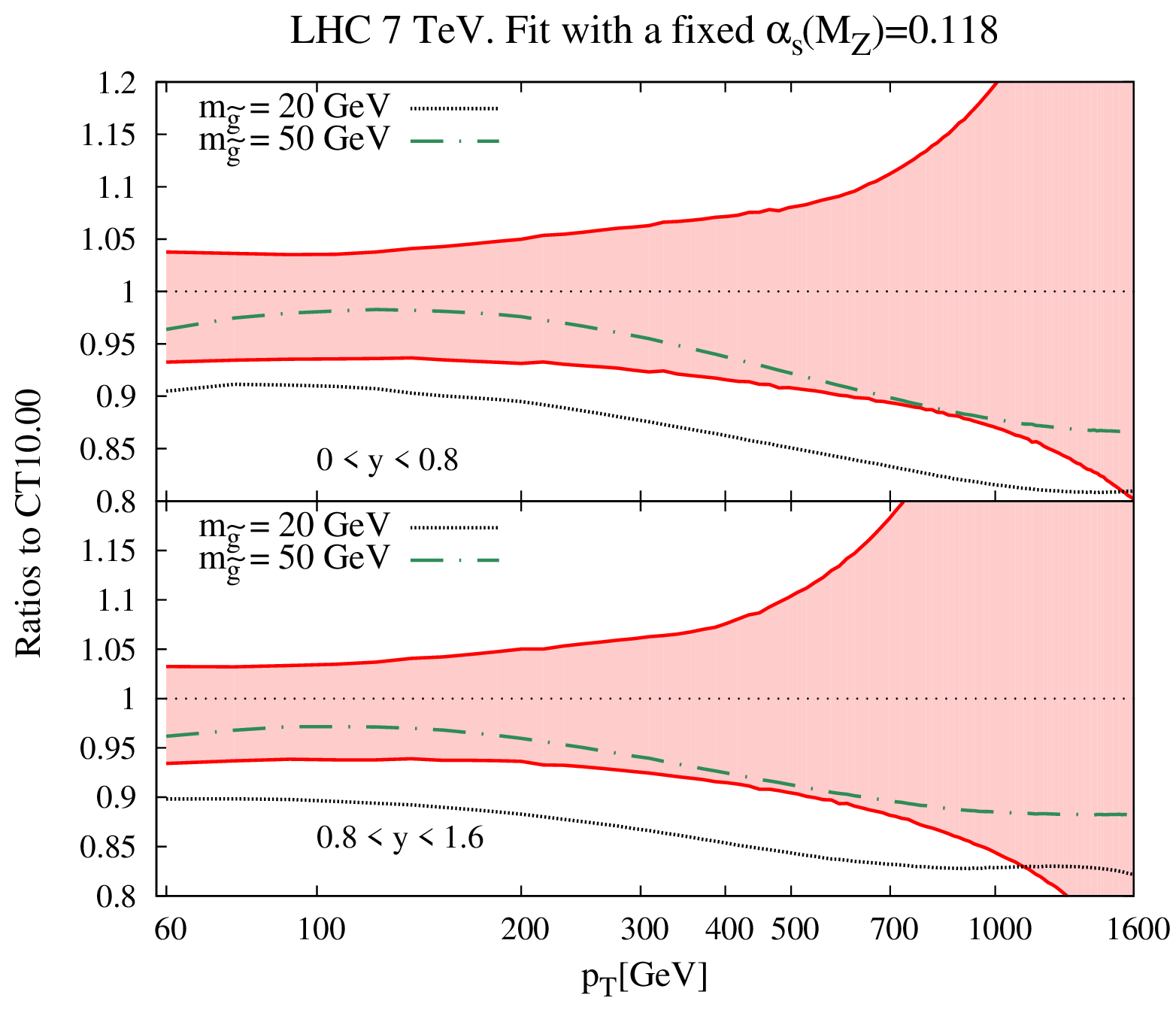}
\includegraphics[width=11cm]{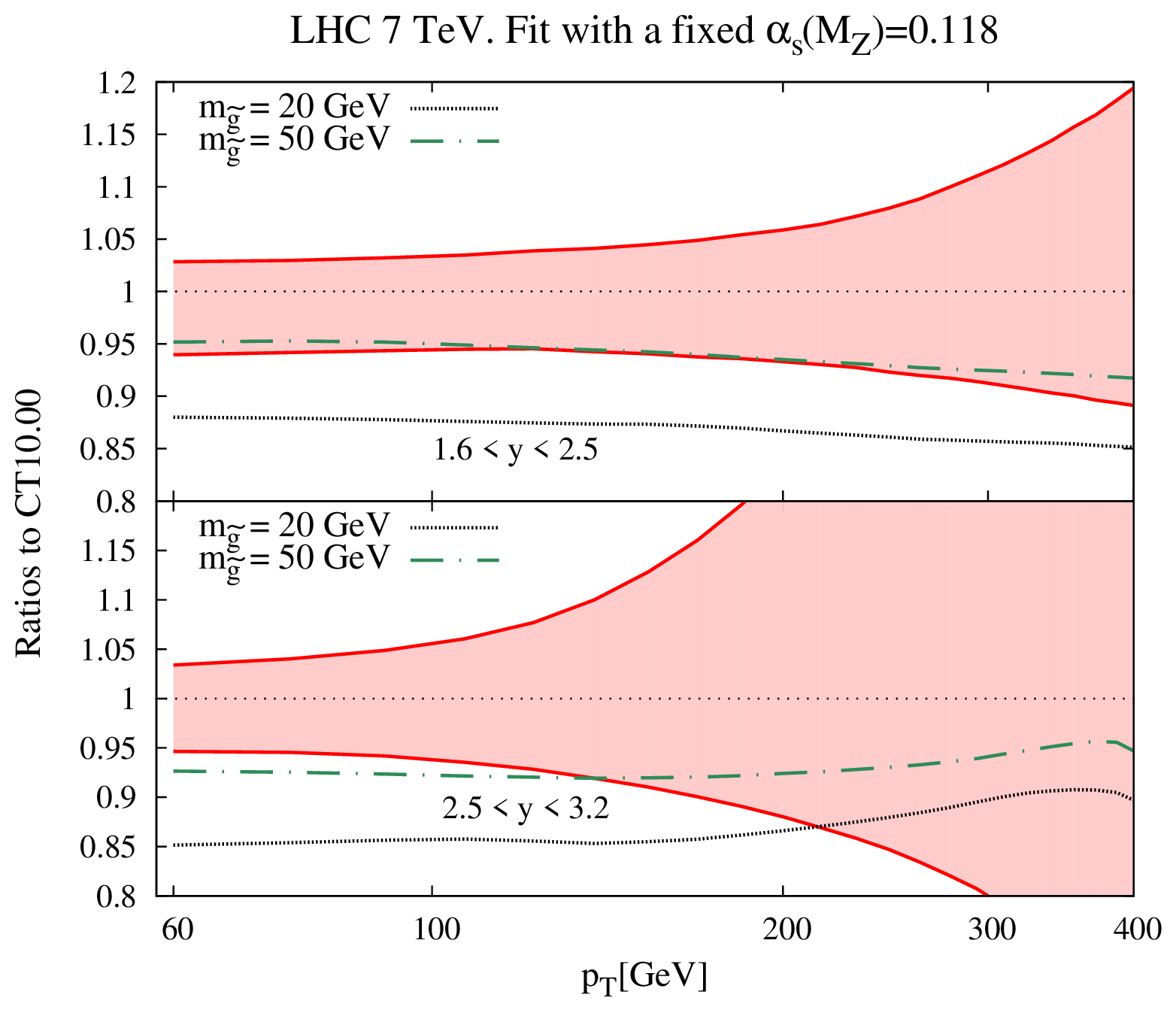}
\caption{\small{Ratios of single-inclusive jet cross sections
at $\sqrt{s}=7$~TeV, obtained
  from the central PDF set of CT10, CT10.00, and the SM+SUSY PDFs for
gluino masses
$m_{{\tilde g}}=20$ GeV (dashed line) and 50 GeV (dot-dashed
     line). The asymmetric PDF uncertainty of the CT10 set is also
     shown as a filled band. The SM+SUSY PDFs are obtained under the
     assumption of  $\alpha_s(M_Z)=0.118$ for both sets.} }
\label{Flhc7}
\end{center}
\end{figure}

\begin{figure}[tp]
\begin{center}
\includegraphics[width=11cm]{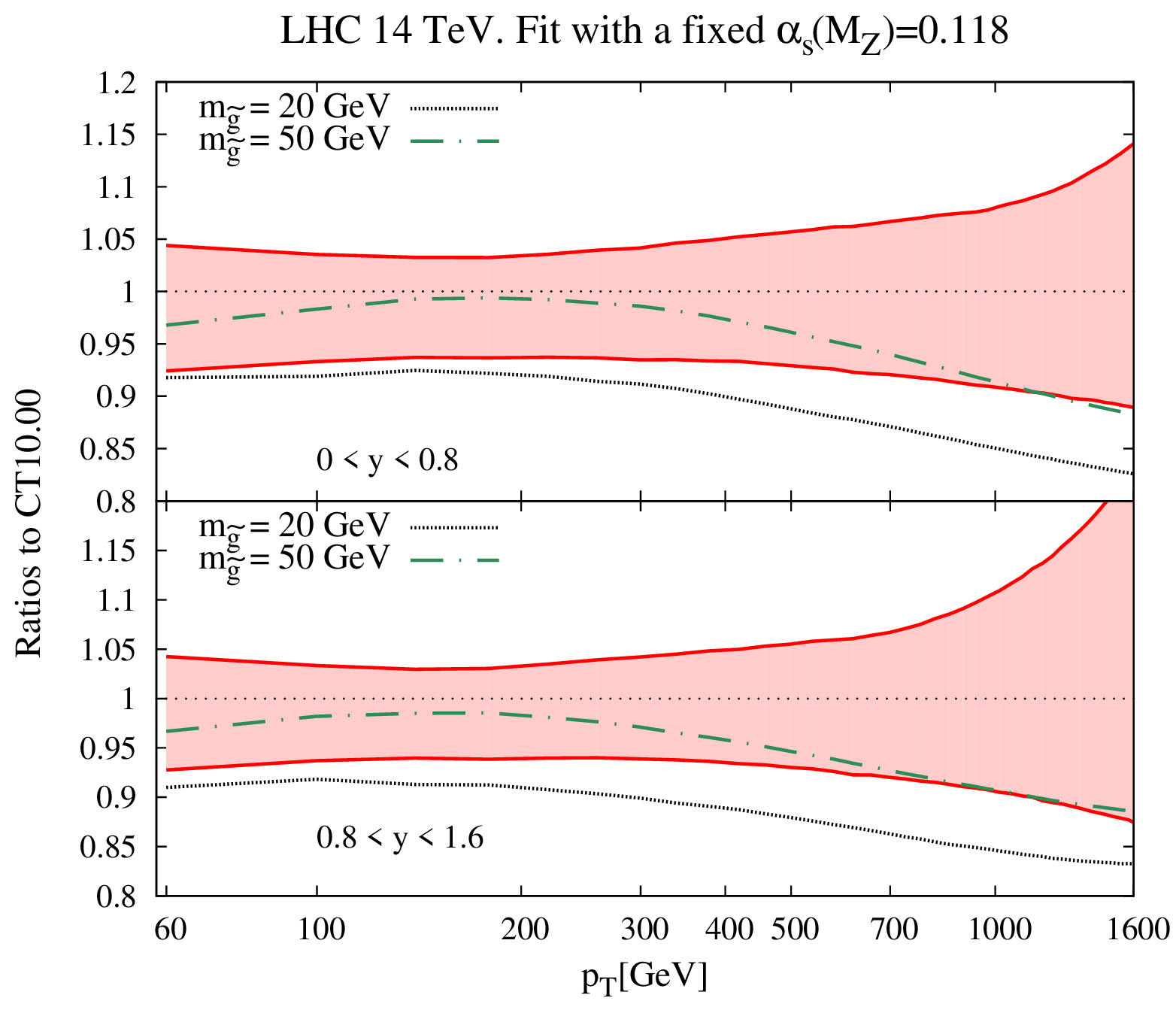}
\includegraphics[width=11cm]{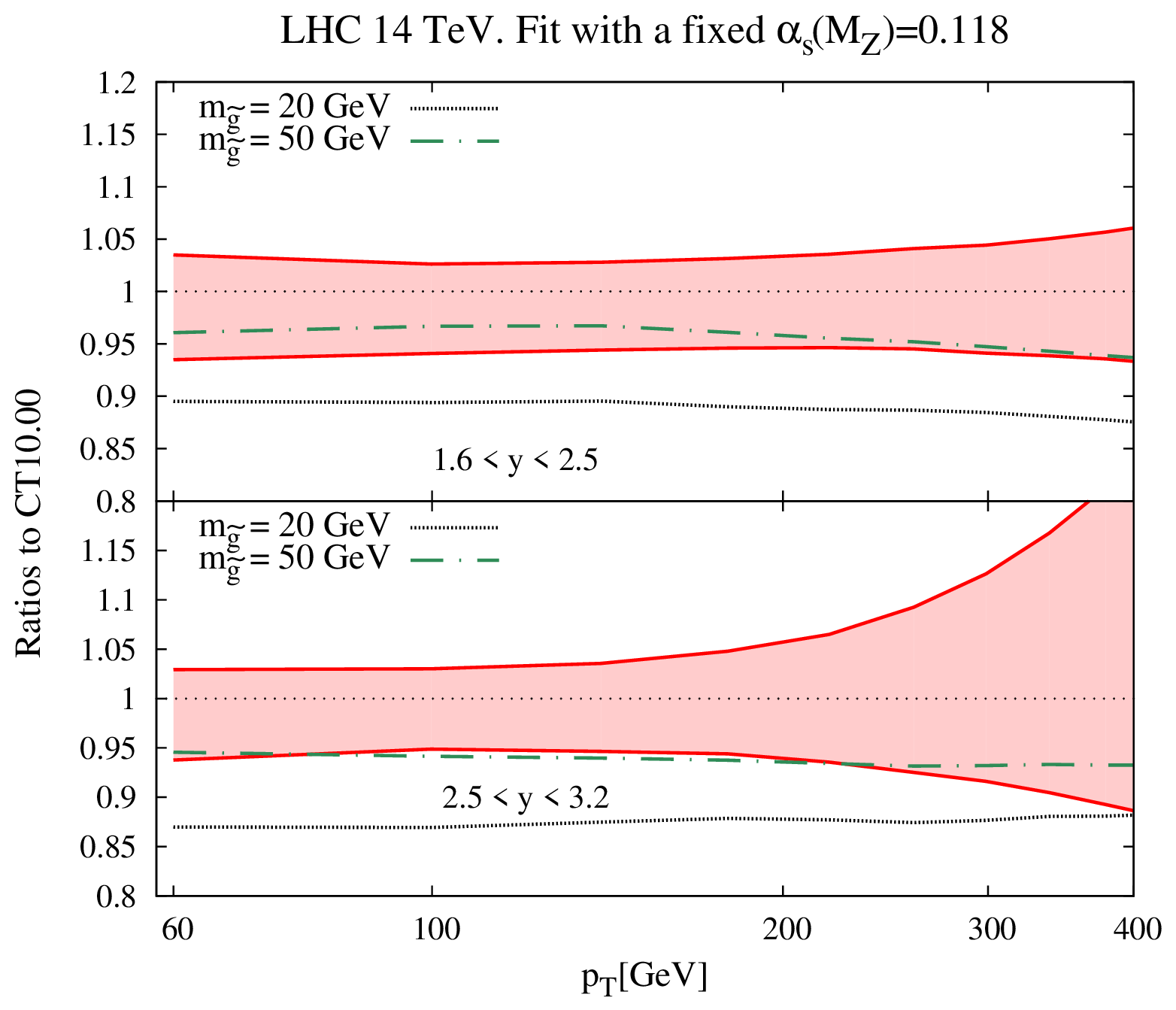}
\caption{\small{Same as Fig.~\ref{Flhc7}, for $\sqrt{s}=14$ TeV.} }
\label{Flhc14}
\end{center}
\end{figure}

\begin{figure}[tp]
\begin{center}
\includegraphics[width=11cm]{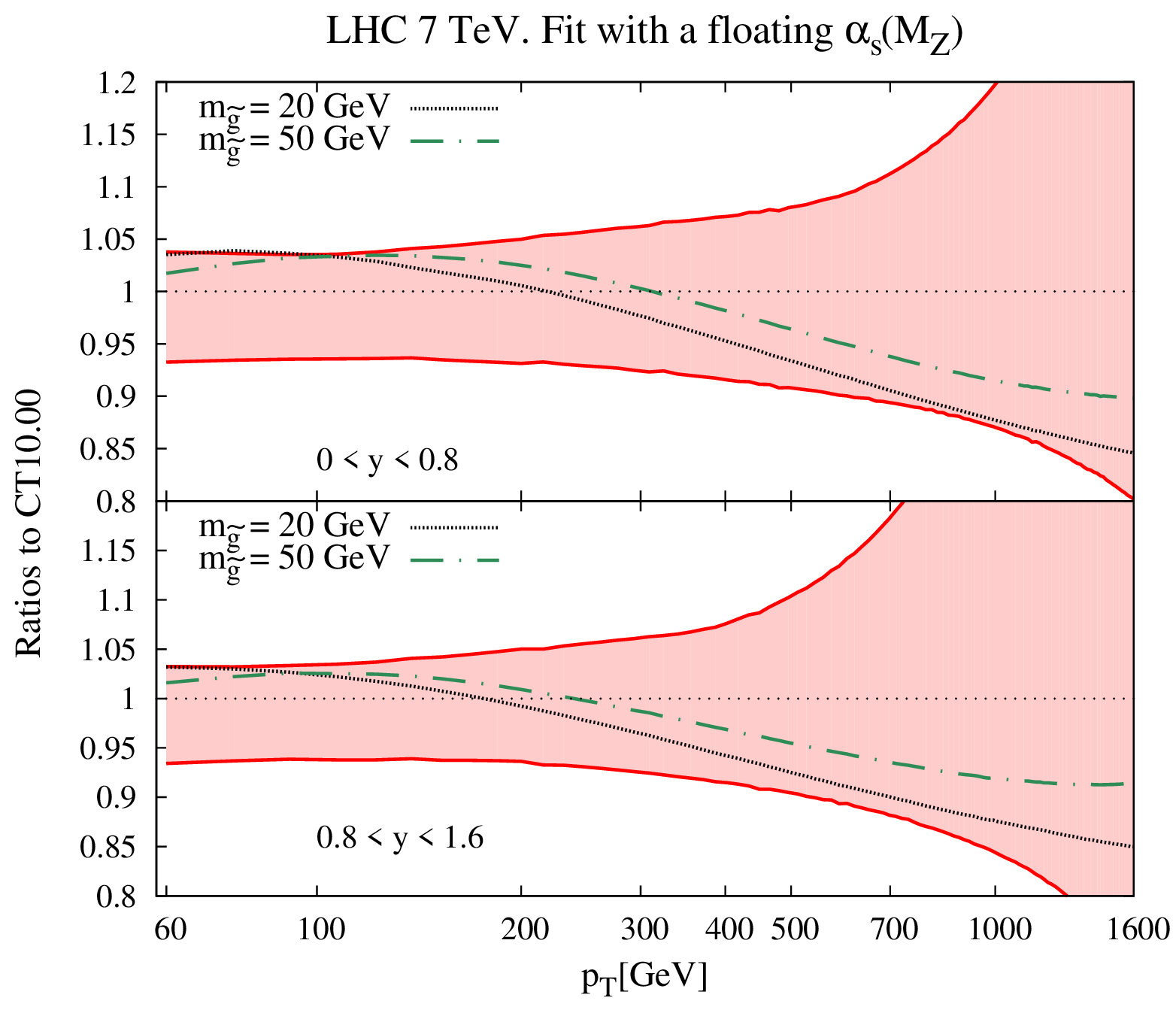}
\includegraphics[width=11cm]{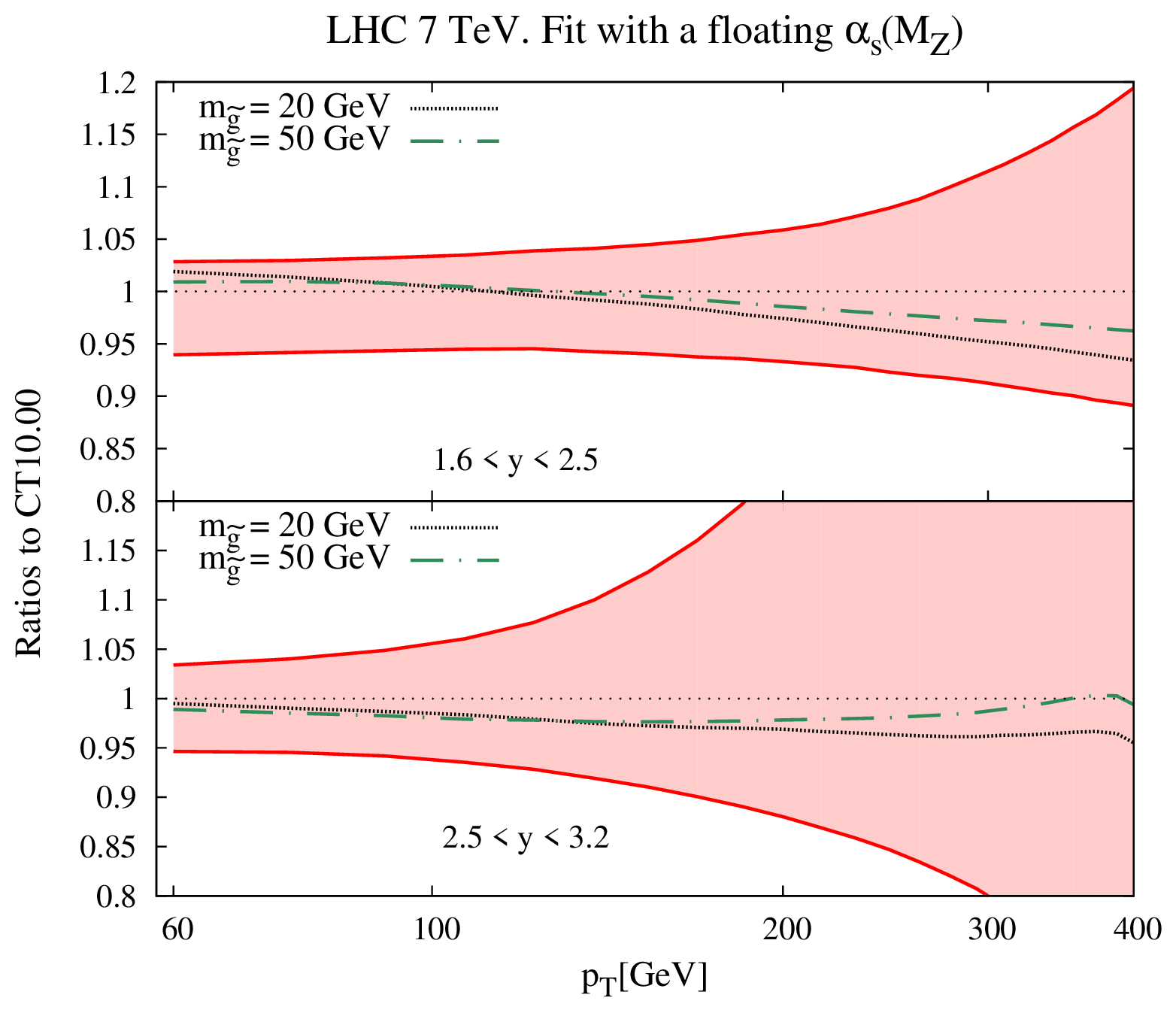}
\caption{\small{Same as  Fig.~\ref{Flhc7}, but with
     $\alpha_s(M_Z)=0.126$ in the SM+SUSY calculation with $m_{\tilde
     g}=20\mbox{ GeV}$,
and  $\alpha_s(M_Z)=0.121$ in the SM+SUSY calculation with $m_{\tilde
     g}=50\mbox{ GeV}$.} }
\label{Nlhc7}
\end{center}
\end{figure}

\begin{figure}[tp]
\begin{center}
\includegraphics[width=11cm]{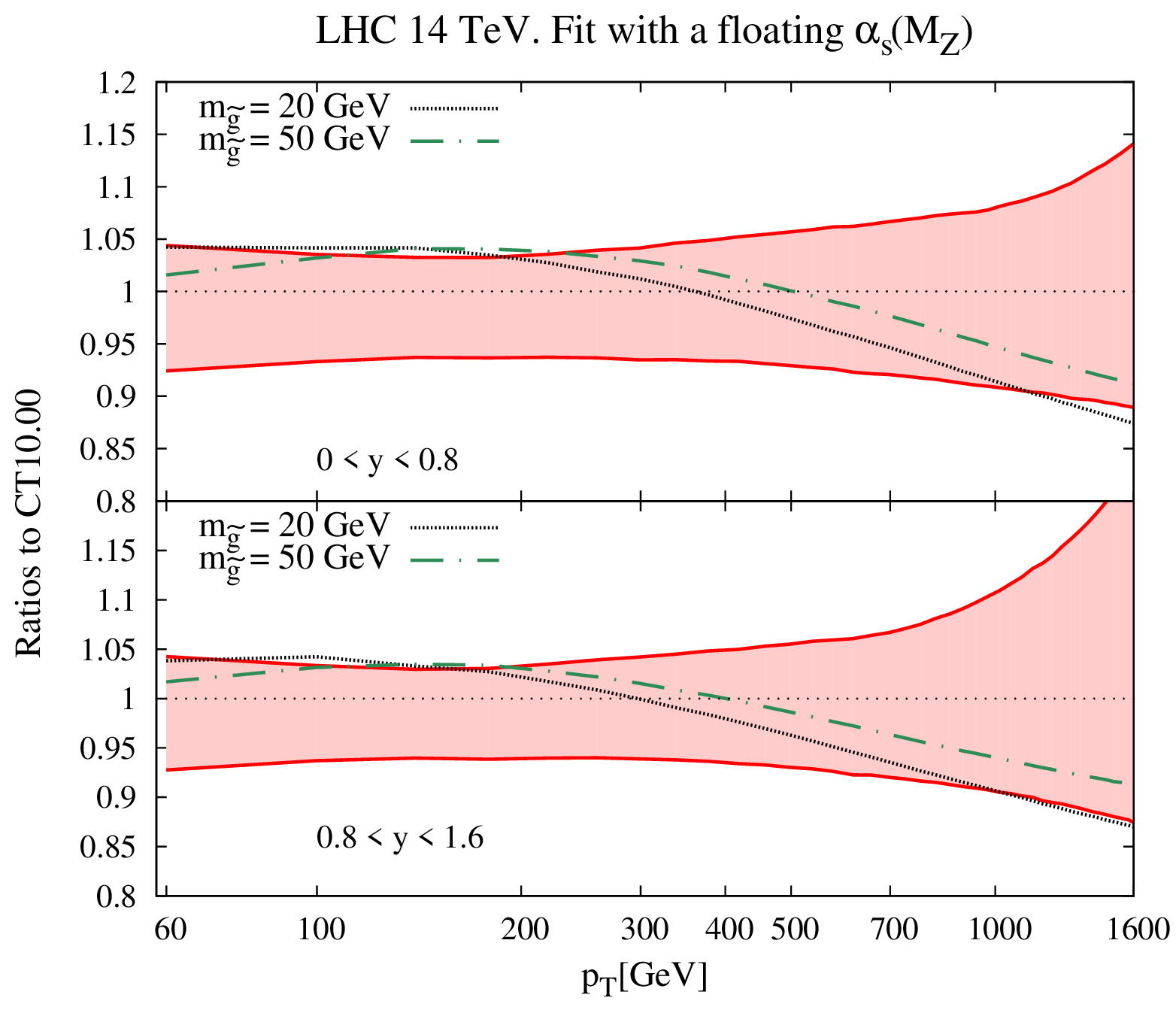}
\includegraphics[width=11cm]{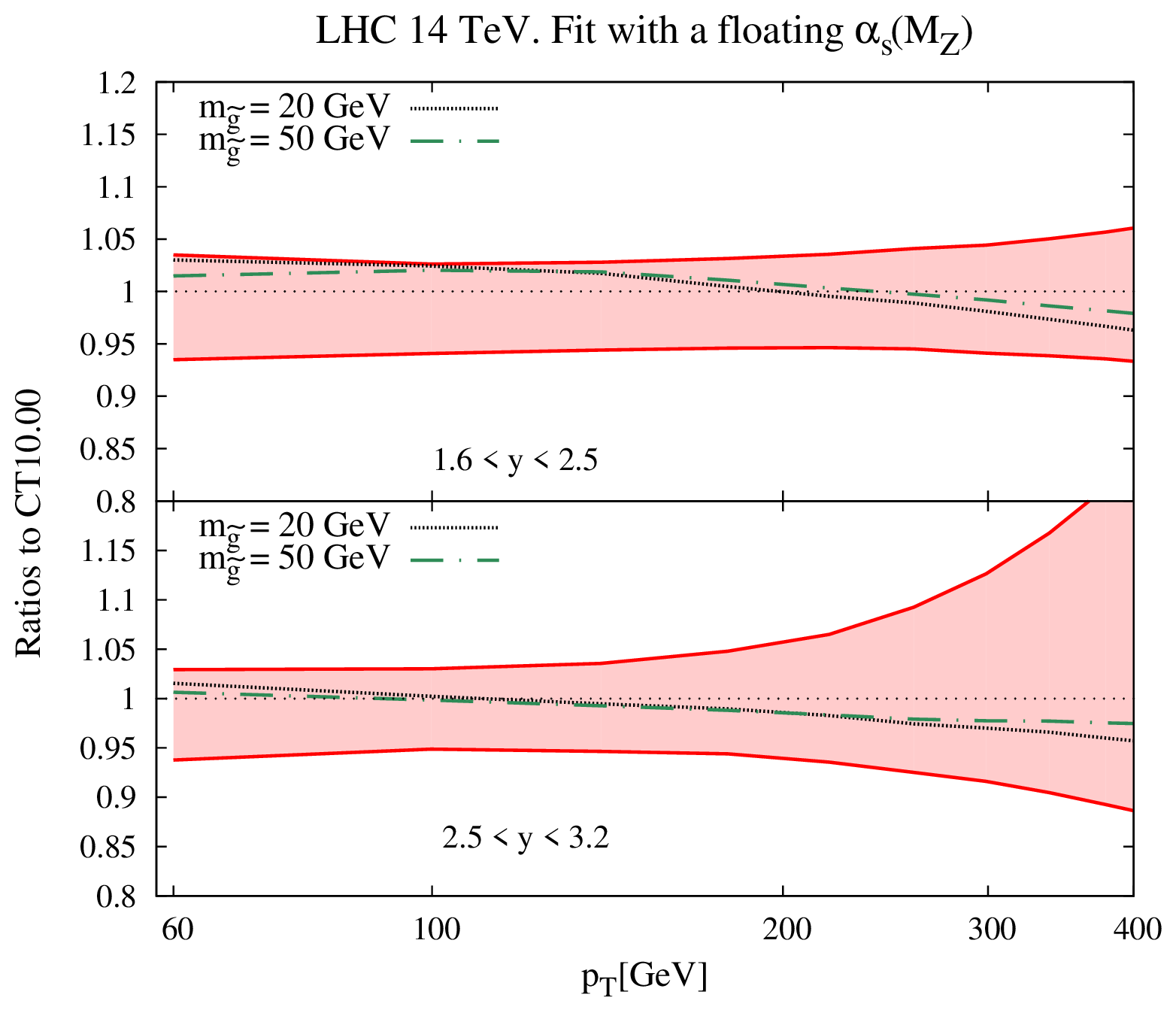}
\caption{\small{Same as Fig.~\ref{Nlhc7}, for $\sqrt{s}=14$ TeV.} }
\label{Nlhc14}
\end{center}
\end{figure}

Compare, for example, single-inclusive jet cross sections at the LHC
energies $\sqrt{s}$=7 and $\sqrt{s}$=14 TeV, 
computed at NLO with the EKS code~\cite{Ellis:1990ek, Ellis:1992en} 
in the pure SM case and in the presence of light gluinos. 
The CT10 asymmetric PDF error bands on the cross sections, normalized
to the predictions based on the central CT10.00 PDF set,
are also shown in Figs.~\ref{Flhc7}-\ref{Nlhc14} as a function of the jet's
transverse momentum $p_T$, in several bins of the jet rapidity $y$. 
Ratios of the expectations based on the SM+SUSY PDFs for
$m_{\tilde g}=20$ and 50 GeV to their counterparts based on
the CT10.00 set are shown as the dashed and dot-dashed lines,
respectively.

In Figs.~\ref{Flhc7} and \ref{Flhc14}, these ratios are computed with 
$\alpha_s(M_Z)=0.118$ assumed in all PDFs and cross sections.
In this case,  the SM+SUSY curves lie outside the respective 
CT10 PDF uncertainty bands for some $p_T$, suggesting that
the SM and SM+SUSY scenarios can be distinguished, 
if sufficient experimental accuracy is achieved.  
On the other hand, if
$\alpha_s(M_Z)$ takes the values of 0.126 and 0.121 that are preferred
in the SM+SUSY fits with $m_{\tilde g}=20$ and 50 GeV, respectively, then the 
SM+SUSY curves lie within the CT10 PDF error bands, as shown in
Figs.~\ref{Nlhc7} and \ref{Nlhc14}.  In this case, discrimination
of the SM and the SM+SUSY cases is more challenging, as reduction of 
the experimental uncertainty below the current PDF uncertainty would be 
necessary.

For the inclusive jet cross sections to provide a good discrimination 
between the SM and SM+SUSY scenarios, the uncertainties on both 
$\alpha_s$ and PDFs must be reduced below the current values.   
NNLO contributions to SM processes and NLO gluino
contributions must also be implemented in both the PDFs and
jet cross sections. 

A different approach to detecting the presence of new colored states could 
be based on the expectation that QCD radiation off a heavy colored object differs 
from that from massless partons that
dominate the inclusive cross sections. It may be possible to identify 
jets containing gluinos by studying distributions in the jet
mass or other jet shapes. The distribution in the jet mass produced by
conventional QCD radiation decreases smoothly as the jet mass
increases. Decays of gluinos would result in jets whose mass
distributions peak at $m_{\tilde g}$, and gluino jet contributions could 
be identifiable above the continuous SM background in the
distributions in the jet mass or related observables, using methods being 
developed~\cite{Thaler:2008ju,Almeida:2008tp,Almeida:2008yp,Ellis:2009su}. 

\clearpage 

\section{Summary and Discussion\label{sec:Conclusion}}
In this paper, we explore modifications in QCD
scattering cross sections introduced by color-octet
Majorana fermions in supersymmetry (gluinos) 
and other popular extensions
of the standard model.  Their influence must be included in
the evolution of the strong coupling strength and the parton distribution
functions, especially if these fermions have mass below 
100 GeV (possible in the absence of model-specific
assumptions).   In addition to modifying the evolution of $\alpha_{s}(Q)$ and 
the PDFs of the SM quarks and gluons, a relatively light gluino also 
introduces new production channels such as $gg\to{\tilde g}{\tilde g}$
in the inclusive jet production case.  In this context, hadronic scattering
data included in global PDF analyses can provide model-independent
constraints on the color-octet particles. 

We examine the values of  $\chi^2$  obtained from our global fits as a 
function of the gluino mass $m_{\tilde g}$.  By analyzing a combination of 
the latest HERA and Tevatron data on hadronic
scattering, and world measurements of the QCD coupling at $Q< 10 $ GeV
and $Q=M_Z$, we conclude that gluinos must be heavier than 25 GeV
at 90\% C.L., if $\alpha_s(M_Z)=0.118$, and heavier than 15 GeV if $\alpha_s(M_Z)$ is
arbitrary.   These constraints supersede the 2004 study based on the CTEQ6 data set,
in which we found a  lower limit on the gluino mass of $m_{\tilde g} > 12$ GeV for  
$\alpha_s(M_Z)=0.118$, and no limit if $\alpha_s(M_Z)$ is
arbitrary \cite{Berger:2004mj}. These new bounds are comparable to the 
gluino mass bounds  $m_{\tilde{g}} > 26.9 $ and 51 GeV obtained from
the analysis of event shapes in $e^+e^-$ hadroproduction at LEP
\cite{Heister:2003hc,Kaplan:2008pt}.   
Our constraints on $m_{\tilde g}$ are 
obtained from the analysis of inclusive QCD observables and are not
affected by theoretical uncertainties of the kind that arise in the determination 
of $\alpha_s(M_Z)$ from the LEP 
data~\cite{Becher:2008cf, Davison:2008vx, Dissertori:2009ik, Abbate:2010vw, Abbate:2010xh} 
and LEP event shapes. 

The changes in $\alpha_s(M_Z)$ and in the PDFs of standard model partons 
must be taken into consideration when QCD tests are made with LHC data.  
The high energy of the LHC and the extended range in jet transverse 
momentum offers hope that BSM deviations from pure QCD will show up in 
inclusive jet cross sections.  As discussed in our comparisons with Tevatron 
jet data, it will be critical to control experimental uncertainties on the jet energy 
scale and jet energy resolution.   Gluino contributions and adjustments in the 
SM parameters tend to offset one another.  The power of precise measurements 
of the LHC single-inclusive jet cross sections will be enhanced provided that 
$\alpha_s(M_Z)$ and the PDFs for gluons and quarks are constrained more
tightly than now by measurements in other channels.  

For the purpose of studying jet properties in detail, we provide
routines to interface with the SM+gluino PDFs. These are linked from
the CTEQ webpage at {\tt cteq.org}.  We also note that the
MadGraph/MadEvent programs \cite{Alwall:2007st} provide a mechanism to
incorporate SUSY PDFs in the initial state; information for using this
interface is also provided on the webpage.

\begin{acknowledgments}
We thank Tom Rizzo for discussions regarding searches for new
physics and bounds on gluino masses, and Mike Whalley and Andy Buckley for
extending the LHAPDF library \cite{Bourilkov:2006cj,Whalley:2005nh}  
to incorporate PDFs with light gluinos.  The authors also thank CTEQ 
members for helpful discussions.  
E.~L.~B. is supported by the U.S. Department of Energy under Grant
No.~DE-AC02-06CH11357. The work at SMU is supported in part
by U.S. DOE contract DE-FG02-04ER41299; 
the U.S. DOE Early Career Research Award DE-SC0003870; 
LHC Theory Initiative Travel Fellowship 
awarded by the U.S. National Science Foundation under grant PHY-0705862;
and by Lightner-Sams Foundation.
H.-L.~L. is supported by the U.S. National Science Foundation 
under grant PHY-0855561 and by National Science Council of Taiwan under grants
NSC-98-2112-M-133-002-MY3 and NSC-99-2918-I-133-001.
E.~L.~B. and P.~M.~N. thank the Aspen Center for Physics 
for hospitality during the summer of 2010 when part of this work was done.  
F.~I.~O thanks CERN for hospitality where a portion of this work was performed.
 
\end{acknowledgments}

\appendix
\section{Modification of the Strong Coupling\label{app:alphas}}

The running of $\alpha_{s}(Q)$ must be matched to the individual
PDF set with the appropriate mass thresholds. The expansion of the
evolution equation for $\alpha_{s}(Q)$ 
\begin{eqnarray}
Q \frac{\partial }{\partial Q }\, \, \alpha_{s}(Q ) & = & -\frac{\alpha_{s}^{2}}{2\pi }\sum _{n=0}^{\infty }\beta _{n}\left(\frac{\alpha_{s}}{4\pi }\right)^{n}
\nonumber \\
 & = & -\left[\beta _{0}\frac{\alpha_{s}^{2}}{2\pi }+\beta _{1}\frac{\alpha_{s}^{3}}{2^{3}\pi ^{2}}+...\right],
\end{eqnarray}
can be solved perturbatively.
It takes the form~\cite{Yao:2006px} \begin{eqnarray}
\alpha_{s}(Q) & = & \frac{4\pi}{\beta_{0}\ln\left(\frac{Q^{2}}{\Lambda^{2}}\right)}\left[1-\frac{\beta_{1}}{\beta_{0}^{2}}\frac{\ln[\ln(Q^{2}/\Lambda^{2})]}{\ln(Q^{2}/\Lambda^{2})}+\right.\nonumber \\
 & + & \left.\frac{\beta_{1}^{2}}{\beta_{0}^{4}\ln^{2}(Q^{2}/\Lambda^{2})}+....\right] .\label{eq:AlphaSEvolution}\end{eqnarray}
The beta functions, $\beta_{0}$ and $\beta_{1}$ depend on the number
of active fermions and bosons. When supersymmetric particles are included~\cite{Machacek:1983tz},
the first two coefficients in Eq.~(\ref{eq:AlphaSEvolution}) are
\begin{eqnarray*}
\beta_{0} & = & 11-\frac{2}{3}n_{f}-2n_{{\tilde g}}-\frac{1}{6}n_{\tilde{f}},\end{eqnarray*}
 and \begin{eqnarray}
\beta_{1} & = & 102-\frac{38}{3}n_{f}-48n_{{\tilde g}}-\frac{11}{3}n_{\tilde{f}}+\frac{13}{3}n_{{\tilde g}}n_{\tilde{f}},\label{beta1}\end{eqnarray}
 where $n_{f}$ is the number of quark flavors, $n_{{\tilde g}}$ is
the number of gluinos, and $n_{\tilde{f}}$ is the number of squark
flavors. As the evolution proceeds 
across mass thresholds, these numbers and, consequently
$\alpha_{s}$, must be adjusted.

\section{Gluino contributions to the single-inclusive jet cross
section \label{app:jet}}

The leading-order cross section for inclusive (di)jet production, 
$H_1 H_2 \rightarrow j(p_3) j(p_4)
 X$, expressed in terms of the transverse momentum $p_T$ and
 rapidities $y_3$, $y_4$ of the jets, is 
\ba
&&\frac{d\sigma}{dp_{T} dy_3 dy_4}=\frac{2\pi\alpha_s^2 ~p_{T}}{\hat{s}^2} 
\sum_{i,j}x_1 x_2 \,f_{H_1\rightarrow i}(x_1,\mu_F^2)\, f_{H_2\rightarrow j}(x_2,\mu_F^2)\,
\sum_{spin}\left| {\mathcal M}_{p_1 p_2 \rightarrow p_3 p_4}  \right|^2\,,
\ea
where $x_1=m_T/\sqrt{s}(e^{y_3} + e^{y_4})$, 
and $x_2=m_T/\sqrt{s}(e^{-y_3} + e^{-y_4})$ are the parton momentum fractions; 
$m_T^2=p_T^2 + m_{{\tilde g}}^2$ is the gluino's transverse mass; and 
$\sqrt{s}$ is the collider center-of-mass energy.
In our analysis, scattering amplitudes for subprocesses with gluino pair
production,  $g g \rightarrow {\tilde g} {\tilde g}$
and $q \bar{q} \rightarrow {\tilde g} {\tilde g}$, are included with
full dependence on gluino mass $m_{\tilde g}$. Scattering amplitudes
for the other LO subprocesses (with at least one initial-state gluino)
are evaluated in the $m_{\tilde g}=0$ approximation, in accord with
the S-ACOT factorization scheme~\cite{Collins:1998rz,Kramer:2000hn}.

SUSY contributions with full mass dependence can be found in the literature (e.g., in
\cite{Beenakker:1996ch} and \cite{Amsler:2008zzb}), but they are presented
here in a consistent notation for completeness. 
In terms of the usual parton-level Mandelstam variables, 
$\hat{s}=(p_1 + p_2)^2$, $\hat{t}=(p_1 - p_3)^2$, and $\hat{u}=(p_1 -
p_4)^2$, the square of the amplitude for $q\bar{q}\rightarrow{\tilde g} {\tilde g}$ is 
\ba
&&\vert {\mathcal M}_{q\bar{q}\rightarrow {\tilde g} {\tilde g} }\vert^2=
\frac{8}{9} \left[
\frac{\hat{s} m_{{\tilde g}}^2}{3 (m_{\tilde{q}}^2-\hat{t})(m_{\tilde{q}}^2-\hat{u})}
+\frac{4 (m_{{\tilde g}}^2-\hat{t})^2}{3(m_{\tilde{q}}^2-\hat{t})^2}
-\frac{3 \left(\hat{s} m_{{\tilde g}}^2+(m_{{\tilde g}}^2-\hat{t})^2\right)}{\hat{s} (m_{\tilde{q}}^2-\hat{t})}
+\frac{4 (m_{{\tilde g}}^2-\hat{u})^2}{3 (m_{\tilde{q}}^2-\hat{u})^2}
\right.\nonumber\\
&&\hspace{3cm}\left.
-\frac{3 \left(\hat{s}m_{{\tilde g}}^2+(m_{{\tilde g}}^2-\hat{u})^2\right)}{\hat{s} (m_{\tilde{q}}^2-\hat{u})}
+\frac{3 \left(2\hat{s} m_{{\tilde g}}^2 + (m_{{\tilde g}}^2-\hat{t})^2+(m_{{\tilde g}}^2-\hat{u})^2\right)}{\hat{s}^2}
\right].
\label{qqb_ginogino}
\ea
Here $m_{\tilde{q}}$ is the mass of the squark, and the pre-factor
$8/9$ is a color factor. We report the expression with all the fermion mass dependence,
but in our computations we have taken the limit
$m_{\tilde{q}}\rightarrow \infty$.

The square of the amplitude for $g g\rightarrow{\tilde g} {\tilde g}$ is
\ba
&&\vert {\mathcal M}_{gg\rightarrow {\tilde g} {\tilde g} }\vert^2=
-\frac{9 m_{{\tilde g}}^6}{4 \hat{s}^2
\left(\hat{t}-m_{{\tilde g}}^2\right)}
-\frac{9 m_{{\tilde g}}^6}{4 \hat{s}^2
\left(\hat{u}-m_{{\tilde g}}^2\right)}
+\frac{27\hat{u} m_{{\tilde g}}^4}{4 \hat{s}^2
\left(\hat{t}-m_{{\tilde g}}^2\right)}
\nonumber\\
&&
-\frac{45
   m_{{\tilde g}}^4}{2 \hat{s}
   \left(\hat{t}-m_{{\tilde g}}^2\right)}
+\frac{27
   \hat{t} m_{{\tilde g}}^4}{4 \hat{s}^2
   \left(\hat{u}-m_{{\tilde g}}^2\right)}
-\frac{45
   m_{{\tilde g}}^4}{2 \hat{s}
   \left(\hat{u}-m_{{\tilde g}}^2\right)}
\nonumber\\
&&
+\frac{27
   m_{{\tilde g}}^4}{\left(\hat{t}-m_{{\tilde g}}^2\right)
   \left(\hat{u}-m_{{\tilde g}}^2\right)}
+\frac{9
   m_{{\tilde g}}^4}{\hat{s}^2}-\frac{81
   m_{{\tilde g}}^4}{\left(\hat{t}-m_{{\tilde g}}^2\right)
   ^2}
-\frac{81
   m_{{\tilde g}}^4}{\left(\hat{u}-m_{{\tilde g}}^2\right)
   ^2}
\nonumber\\
&&
-\frac{27 \hat{u}^2 m_{{\tilde g}}^2}{4
   \hat{s}^2
   \left(\hat{t}-m_{{\tilde g}}^2\right)}
-\frac{9
   \hat{t} m_{{\tilde g}}^2}{\hat{s}^2}
+\frac{45 \hat{u} m_{{\tilde g}}^2}{2 \hat{s}
   \left(\hat{t}-m_{{\tilde g}}^2\right)}
-\frac{9 \hat{u} m_{{\tilde g}}^2}{\hat{s}^2}
+\frac{9 m_{{\tilde g}}^2}{\hat{s}}
\nonumber\\
&&
-\frac{27 \hat{t}^2
   m_{{\tilde g}}^2}{4 \hat{s}^2
   \left(\hat{u}-m_{{\tilde g}}^2\right)}+\frac{45
   \hat{t} m_{{\tilde g}}^2}{2 \hat{s}
   \left(\hat{u}-m_{{\tilde g}}^2\right)}
+\frac{9
   \hat{t}^2}{4 \hat{s}^2}
+\frac{9
   \hat{u}^2}{4 \hat{s}^2}
+\frac{9 \hat{t}
   \hat{u}}{2 \hat{s}^2}
\nonumber\\
&&
+\frac{9
   \hat{u}^3}{4 \hat{s}^2
   \left(\hat{t}-m_{{\tilde g}}^2\right)}
+\frac{9 \hat{t}^3}{4
   \hat{s}^2 \left(\hat{u}-m_{{\tilde g}}^2\right)}
\label{gg_ginogino}.
\ea

\section{Parton Luminosities  \label{app:lumi}}

Parton-parton luminosity functions portray the relative size of various partonic 
contributions. The parton
luminosity is defined as a convolution integral of the PDFs $f_{i}(\xi,Q)$
for two incoming partons ($i,j={\tilde g},g,u,d,s,...)$: \[
\frac{d{\cal L}_{ij}(\tau,Q)}{d\tau}=f_{i}\,\otimes\, 
f_{j}=\int_{\tau}^{1}\frac{d\xi}{\xi}f_{i}(\xi,Q)\, f_{j}\left(\frac{\tau}{\xi},Q\right),\]
 where $\tau=\hat{s}/s$. Here $\hat{s}$ is the square of the center
of mass energy in the incident parton-parton system. In terms of this
luminosity, the production cross section for a specific reaction is
\begin{equation}
\sigma(s)=\sum_{i,j}\int_{\tau_{0}}^{1}d\tau\,\widehat{\sigma}_{ij}(\tau)\,
\frac{d{\cal L}_{ij}(\tau,Q)}{d\tau}.\label{eq:sigLum}
\end{equation}
The sum is over the initial-state parton flavors $i$ and $j$,
and $\widehat{\sigma}_{ij}(\tau)$ is the partonic cross section for
the subprocess initiated by partons $i,j$.

\begin{figure}[t]
\includegraphics[width=8cm,angle=-90 ]{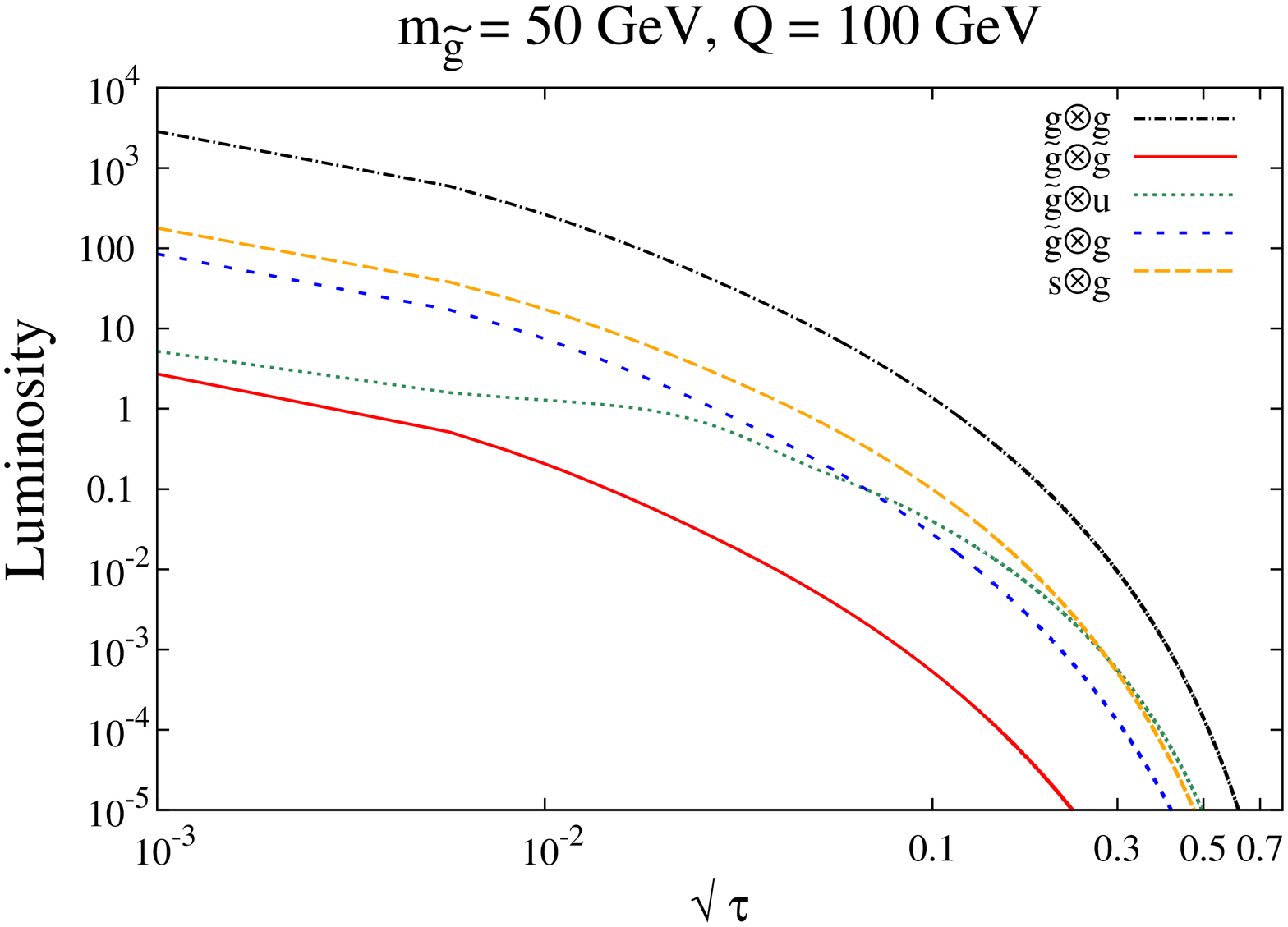}
\includegraphics[width=8cm,angle=-90 ]{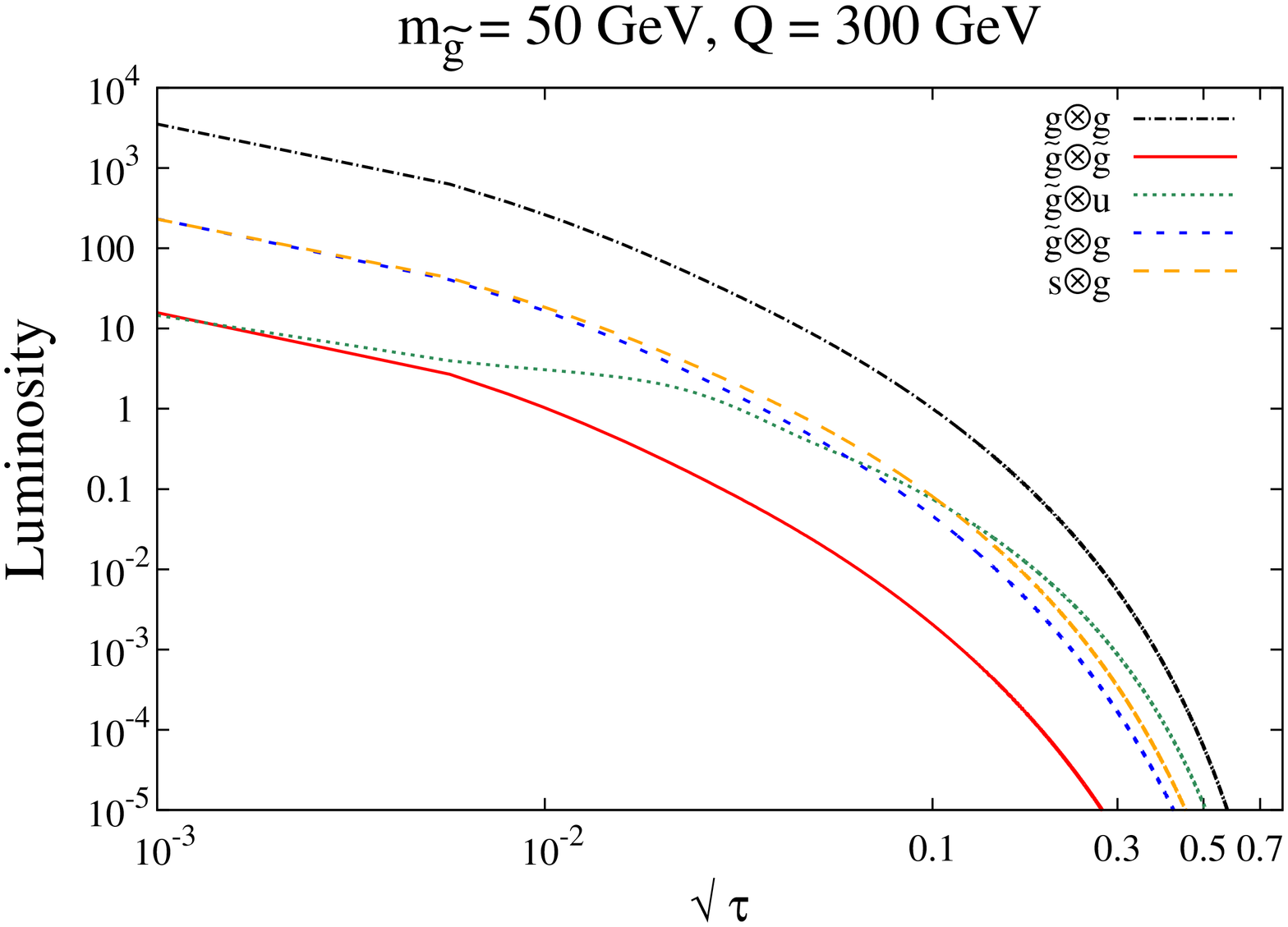}
\caption{\small Parton-parton luminosity $\tau\, d{\cal L}_{ij}(\tau,Q)/d\tau$
vs. $\sqrt{\tau}$ for $m_{{\tilde g}}=50$~GeV at $Q=$100 and 300 GeV.
}
\label{fig:lum}
\end{figure}

The luminosities for some flavor combinations are shown in
Fig.~\ref{fig:lum} for $m_{\tilde g}=50$ GeV.
At $Q=100$ GeV all gluino luminosities are smaller than the SM luminosities, 
but they grow in magnitude as $Q$ increases.  The gluon-gluino luminosity is 
roughly the same as the gluon-bottom quark luminosity, as would be expected 
from the momentum fractions presented in Table I.  At $Q=300$ GeV 
the ${\tilde g}\otimes g$ contribution is comparable to that of the ordinary quarks.  
The ${\tilde g}\otimes g$ combination is smaller than $s\otimes g$ throughout the 
$x$ range for $Q=100$ GeV.   At $Q=300$ GeV, the evolution of the gluino 
is enhanced, and ${\tilde g}\otimes g$ exceeds various SM pairings for $x > 0.1$.   

\providecommand{\noopsort}[1]{}\providecommand{\singleletter}[1]{#1}%

\end{document}